\documentclass[a4paper,11pt]{article}
\pdfoutput=1 

\usepackage{jcappub} 

\usepackage[T1]{fontenc} 

\title{Red Giant evolution in Modified Gravity}

\author[a,1]{Sh. Najafi,\note{Corresponding author.}}
\author[a,1]{M. T. Mirtorabi,}
\author[a]{Z. Ansari,}
\author[b]{and D. F. Mota}

\affiliation[a]{Alzahra University,\\Tehran Province, Tehran, Vanak Village Street, Iran}
\affiliation[b]{Institute of Theoretical Astrophysics,\\ University of Oslo, P.O. Box 1029 Blindern, N-0315 Oslo, Norway}

\emailAdd{sh.najafi@alzahra.ac.ir}
\emailAdd{torabi@alzahra.ac.ir}
\emailAdd{za.ansari@student.alzahra.ac.ir}
\emailAdd{d.f.mota@astro.uio.no}

\abstract{In this paper, we study the chameleon profile in inhomogeneous density distributions and find that the
 fifth force in thin shell near the surface is weaker from what expected in homogeneous density distributions.
Also, we check the validity of quasi-static approximation for the chameleon scalar field in the astrophysical time scales.
 We have investigated the rolling down behavior of the scalar field on its effective potential inside a one solar mass red giant star by using MESA code. We have found that the scalar field is fast enough to follow the minimum of the potential.
 This adiabatic behavior reduces the fifth force and extends the screened regions to lower densities where the field has smaller 
mass and was expected to be unscreened. As a consequence, the star evolution is similar to what expected from standard general relativity.
In addition, considering the stability of star, an approximate constraint on the coupling constant $\beta$ is found.}


\begin{document}
\maketitle
\flushbottom
\section{Introduction}
\label{sec:intro}

The consistency of General Relativity (GR) with observation in cosmological scales requires the existence of some exotic type of fluids with no known 
interaction with normal matter~\cite{1,2,3}. With no observational gravity-independent-evidence for the dark sector, alternatives to Einstein's gravity 
are being put forward as a possible explanation to the Universe acceleration and as alternatives to a dark matter component as well.
 These are generally called theories of modified gravity ~\cite{4,5,6,7,8,9,10,11,12,13,14}.
 The simplest modified gravity (MG) theories are based on imposing 
an extra scalar degree of freedom which might be screened in small scales since GR is well tested in the solar system. Among screening mechanisms, 
chameleon implements a massive scalar field, conformally coupled to metric of the universe. The mass of the scalar field depends on the local 
density of nonrelativistic matter. In high density environments, the field acquires high mass and experiences Yukawa depression and standard GR is 
restored~\cite{16}.

Although in small scales screening prohibits any observational search for evidence of GR modification, there have been hopes that in astrophysical 
scales there might be unscreened regions where extra scalar field acquires gradient and the resulting fifth force could affect stellar
 evolution~\cite{17,18,19,20}.
 In~\cite{21} and~\cite{22}, the modified gravity effect on the main sequence and red giant branch stars in chameleon models
 is considered. It is shown that stars are brighter and hotter in MG compared to GR. Using extrasolar planet datum, new constraints in the 
parameter space of chameleon and symmetron models are 
found in~\cite{23}.
In~\cite{24}, the Chandrasekhar mass limit in Starobinsky f(R) model is studied and found that depending on the modified gravity parameter, the 
limiting mass of white dwarf can be so much smaller or greater than GR which makes it possible to explain super-SNIa and sub-SNIa under unified 
description. The period of stellar oscillation in chameloen model is studied in~\cite{25}. The measurement of TRGB luminosity and Cepheid 
P-L relation within screened and unscreened galaxies is considered as a test of modified gravity in~\cite{26}. 
A new method for constraining 
parameter space of chameleon model is introduced in~\cite{27} taking the differences between gaseous and stellar rotation curves of isolated Dwarf 
galaxies.
Current bounds on parameter space of chameleon and symmetron models from astrophysical, solar system and laboratory tests are presented in a single 
parametrization in review paper~\cite{28}.

	The aim of this work is to find a more accurate implementation of modified gravity on  stellar evolution. 
In previous investigations the equation of motion for the scalar field in and outside of a spherically symmetric star has been solved
 by considering quasi-static limit~\cite{21}. It is assumed that deep inside the star,  density is high enough to put the scalar field on the minimum of the effective potential. 
Depending on density distribution, it is possible to have screened star where field gradient is limited to the difference between minimums 
inside and outside of the star (Thin shell). This assumption might be realistic mainly in main sequence phase since the star does not alter 
its density distribution significantly.
 but at the end of main sequence where the star extends its atmosphere and develops low density regions 
 the field might not be able to reach new minimum by assuming quasi-static approximation. Therefore there would be screening radius beyond which 
the star is unscreened. Because of this unscreened region the stellar evolution might be different from general relativity significantly.

In this paper first, we study the static field profile in homogeneous and inhomogeneous density distributions then
 by direct comparison of scalar field evolution time scale with dynamical time scale of stellar evolution produced by MESA code, we check that whether
 the adiabatic assumption that the scalar field quickly rolls down to its new potential minimum as long as the atmosphere of the star extends and 
develops spherical shell with lower densities. This will lead us to an attractor solution where the value of the field can be obtained directly from
the local density of the spherical shell. Then density profile of the outer atmosphere will resemble the field gradient or evolving fifth force which 
can have impact on further stellar evolution. In the third part, we have studied the stability of nonrelativistic star numerically 
in order to constrain the coupling constant.

\section{Field profile in homogeneous and inhomogeneous density distributions}

\subsection{Field profile in homogeneous density distributions}

\begin{figure}[tbp]
\centerline{\includegraphics[scale=0.42]{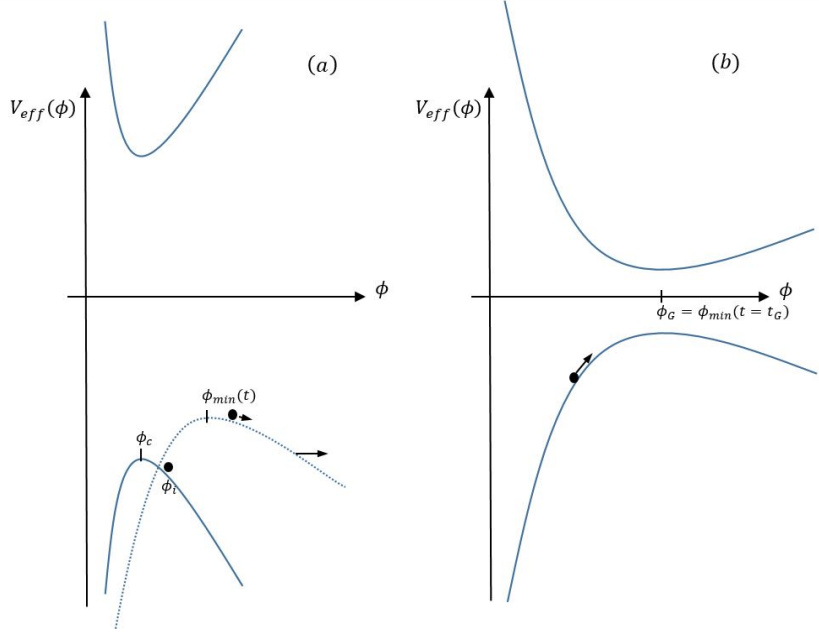}}
\caption{\label{fig:1}Solution of static field equation is the same as rolling down of a particle in an inverted potential
initiated from rest at $\phi_{i}$ , (a, solid curve). In nonhomogeneous densities the particle rolls down
in an evolving hill where its top moves to the right (a, dotted curve). At the end the hill has
overtaken the particle where it must climb up the hill. Both hilltop and particle must meet at
$t = t_{G}$ in $r = r_{G}$ which is equivalent to $\phi = \phi_{G}$ (b).}
\end{figure}

The chameleon field profile inside and outside of a spherical body with homogeneous density $\rho_{c}$
immersed in a homogeneous background density $\rho_{G}$ was derived by Justin Khoury and Amanda
Weltman in their pioneering paper~\cite{16}, which introduced chameleon to the community. They used static
equation of motion
\begin{equation}
\frac{\partial ^{2}\phi}{\partial r^{2}}+\frac{2}{r}\frac{\partial\phi}{\partial r} = \frac{\partial V}{\partial \phi}+ \frac{\beta}{M_{P}}\rho(r)
\end{equation}
and try to solve the equation for a homogeneous density profile such that
\[  \rho(r)=\left\{ \begin{array}{ll}
         \rho_{c} & r< R,\\
         \rho_{G} & r> R.\end{array} \right. \]
where $R$ is radius of spherical body. The potential $V(\phi)$ is assumed to be of the run away form in
such a way that the combined effective potential
\begin{equation}
V_{eff} = V(\phi) + \frac{\beta}{M_{P}}\rho\phi.
\end{equation}
acquires an absolute minimum which induces a mass to the field depends on local environments.
To give an intuition of $\phi(r)$ behavior they convert the boundary value problem, to a
dynamical problem of motion of a particle by assuming $r$ as time and $\phi$ as position of the particle
moving in a potential $-V_{eff}$ and suffering from a dissipative (drag) force $\frac{-2}{t}\frac{dr}{dt}$. By the way the
potential is time dependent since $\rho(r)$ is $r$ dependent. As figure~\ref{fig:1}a shows the dynamical problem
resembles rolling down of the particle from a hill. The hilltop is characterized by minimum of the
potential at center of the sphere, $\phi_{c}$ which is constant up to $t = t_{R} (r = R)$.
Spherical symmetry imposes a boundary condition $\frac{d\phi}{dr} = 0$ at center which might be translated as the particle is initially
 at rest in $t = t_{i} = 0$. In early times the drag force is huge and
the particle must be frozen in its initial position $(r_{i} \equiv \phi_{i} )$. As the time goes on the drag force
reduces and the particle starts to accelerate and gain kinetic energy in response to driving force
$-\frac{dV_{eff}}{d\phi}$
. At $t = t_{R}$ the potential suddenly changes shape and moves to the right such that the
particle finds itself now in a climb up path where it must lose energy to get to the top of the hill
figure~\ref{fig:1}b. Since chameleon is assumed to be relaxed in its minimum of the potential in cosmological
scale $(\phi_{r = r_{G}}  = \phi_{G} )$ the particle must attain to the top of the inverted potential at $t = t_{G}$,
with zero velocity. Then the problem is finding appropriate $r_{i} \equiv \phi_{i}$ where energy gained during
$t < t_{R}$ is exactly equal to energy lost during $t_{R} < t < t_{G}$ .

\subsection{Field profile in inhomogeneous distributions}
In a homogeneous body the particle which resembles field value will accelerate from rest as long
as $t < t_{R}$ which means the chameleon field must deviate from its potential minimum $\phi_{c}$ which is
constant inside the body. In thin shell approximation where $\phi_{i} - \phi_{c} \ll \phi_{c}$ , $\phi$ stays approximately
equal to $\phi_{c}$ up to a very thin shell close to body surface where field starts to acquire gradient and
fifth force is active. At $t = t_{R}$ the minimum at $\phi_{c}$ disappears and a new minimum reappears at $\phi_{G} (\phi_{G} \gg \phi_{c} )$. 
 After this the particle decelerates until finally stop at top of the inverted minimum.

In a nonhomogeneous body, minimum of potential depends on local density, we denote it by
$\phi_{min}(r)$. Since regularly in astrophysical objects $\frac{d\rho}{dr}<0$, we expect that $\phi_{min}(r)$ increases with $r$.
This means the inverted potential maximum (the hilltop) migrates to the right in figure~\ref{fig:1} (dotted
curve). The dynamical problem here is the same as previous one except that as the particle rolls
down the hill, The hilltop transfers to the right. Both the particle and hilltop must arrive at rest
in $\phi_{G}$ at the same time with zero velocity (both field and density are homogeneous in cosmological
scales).
On the ground of what we understand from nature of dynamical problem we can deduce these relations between $\phi_{c}$, $\phi_{i}$ and $\phi_{min}(r)$.

\begin{itemize}
\item  $\phi_{i}>\phi_{c}$:

This is obvious since for $\phi_{i} < \phi_{c}$ the particle would be on the left of the hilltop and rolls down to the
lower value of the field. Because $\phi_{min}(r)$ transfers to right with time the particle never reaches
to $\phi_{G}$ at $t_{G}$ except than $\phi_{i}>\phi_{c}$.

\item $\delta \equiv (\phi_{r}-\phi_{min})  \rightarrow 0 $ at least in early times (inside the star):

Cosmological homogeneity requires

\begin{equation}
\frac{d\phi_{min}}{dr}\lvert_{\phi=\phi_{G}}=\frac{d\phi}{dr}\lvert_{\phi=\phi_{G}}=0.
\end{equation}

This means both hilltop and particle must arrive at $\phi_{G}$ with zero velocity. Then the particle
must experience a phase of deceleration before reaching $\phi_{G}$ . This is not possible except that
the particle shifts to the left of the hilltop some where in between. This requires that inside the star we must have 
$\frac{d\phi_{min}}{dr}>\frac{d\phi}{dr}$; the particle moves slower than the hilltop and $\phi(r)$
lives closer to its minimum than the homogeneous case. There is a point that the field is exactly in its minimum. This is
completely in opposite to what we had in homogeneous sphere. There, since $\phi_{c}$ was fixed and
particle rolls down, the field was deviating more and more from $\phi_{c}$. Here $\phi_{min}(r)$ starts the
race with lower value $(\phi_{c} )$ than the field $(\phi_{i} )$ but it has to overtake the field before reaching
$\phi_{G}$ at the end; then it has to approach to $\phi(r)$ and passes through it.

\item $\delta$ is directly related to $\frac{d\rho(r)}{dr}$:

A smooth density variation is correspond to a slow motion of $\phi_{min}(r)$ to the right. Since
$\phi_{min}(r)$ must overtake $\phi(r)$ before ending the course, $\phi(r)$ must stay close to the hilltop and
experience less driving force and acquire smaller velocity to let $\phi_{min}$ (hilltop) to arrive to the
$\phi_{G}$ before it. This indicates lower $\frac{d\rho(r)}{dr}$ is correspond to lower $\delta$.

Up to now we have ignored the effect of drag force $\frac{2}{r}\frac{d\phi}{dr}$. Following the line of reasoning given
in this note it is clear that this drag will force $\phi(r)$ closer to $\phi_{min}(r)$. The drag force keeps the
particle frozen in very early times and reduces $\delta$ since $\phi_{min}(r)$ moves to the right and experiences no
dragging. This just reduces $\delta$ and never inverts it, because as we stated in first item above, $\phi_{min}(r)$
cannot exceed $\phi(r)$ in very early times since the particle must acquire enough kinetic energy to
climb up the hill at the end of the course.
\end{itemize}

In the pioneering work that introduced the idea of chameleon screening, Khoury and Weltman have solved the equation of motion of 
scalar field inside homogeneous sphere and found that for a high density and large spherical body like Earth, the scalar profile inside the body is 
almost constant and close 
to its minimum of the potential. As long as the field stay constant in high density region the fifth force vanishes and the screening mechanism works.
  Close to the surface of a homogeneous sphere, the field must rise up to match to the outside where the field resides in its cosmic minimum. 
This gradient induces fifth force in outer layers of the sphere. The thin unscreened layer in the surface has two distinctive feature respect to
 screened inside. i) The field deviates from its 
minimum of the potential, ii) for high density and large objects (thin shell approximation) the gradient or fifth force is large. 

\begin{figure}[tbp]
\includegraphics[scale=0.38]{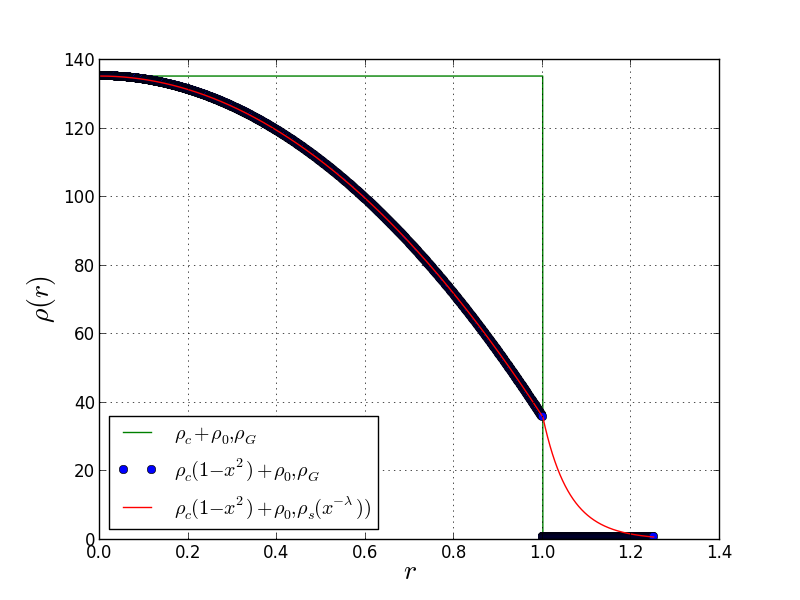}
\includegraphics[scale=0.38]{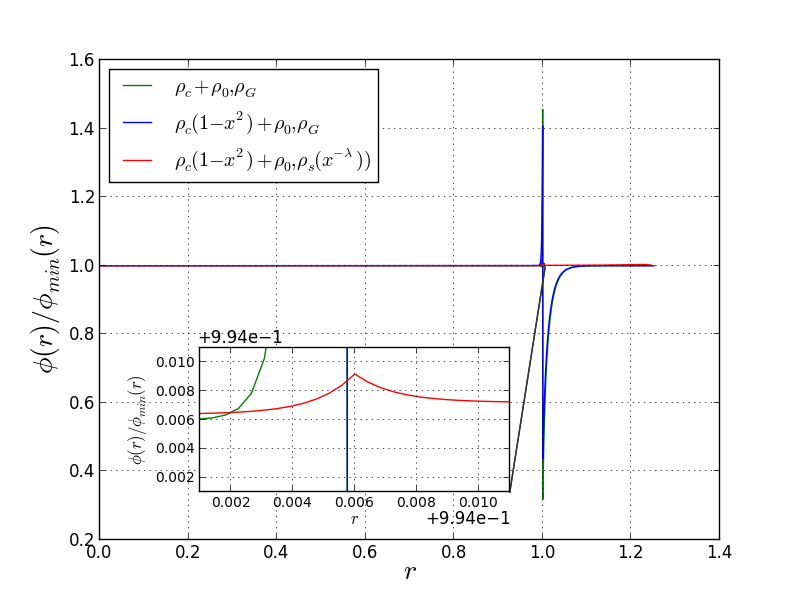}
\includegraphics[scale=0.38]{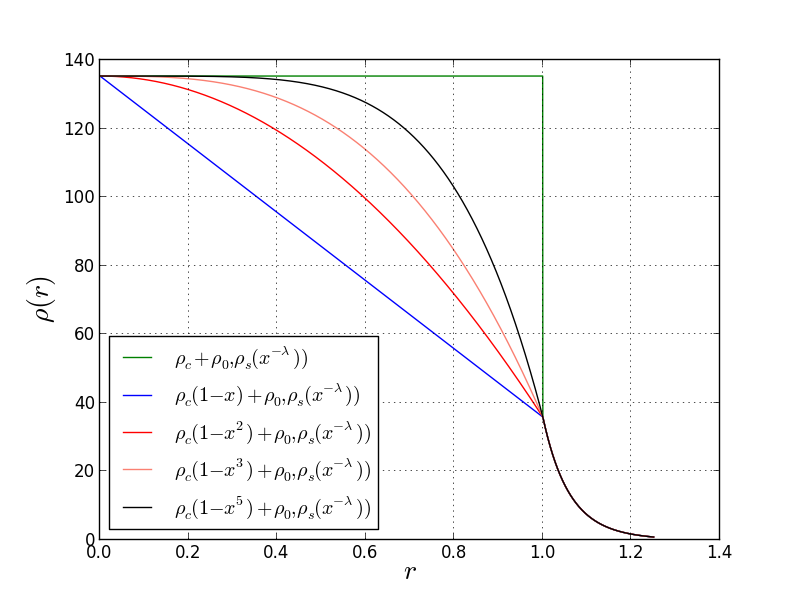}
\includegraphics[scale=0.38]{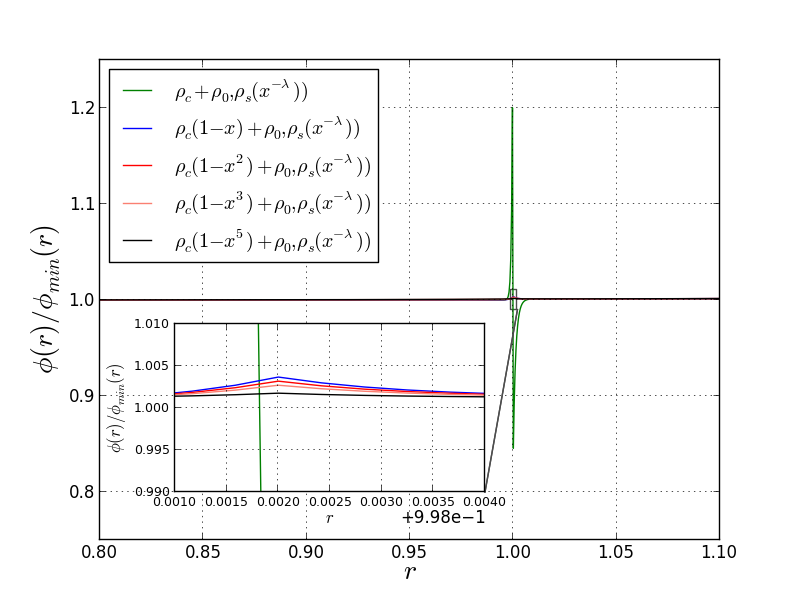}
\caption{\label{fig:inhomo13}Top plots show that the field gradient in thin shell near the surface is smaller when the density is inhomogeneous and connected 
 smoothly to $\rho_{G}$ outside(red line). In the bottom plots: the field gradient in thin shell near the surface is getting smaller choosing 
smoother density profiles.}\label{inhomo13}
\end{figure}

Our basic motivation is to check how this thin unscreened shell varies if we move to nonhomogeneous sphere. 
The static scalar field equation of motion for different density profiles inside and outside of an imaginary model is solved using 
relaxation method \cite{15}. Free parameters are chosen in such a way that there is a thin unscreened shell near the surface. 
In figure~\ref{fig:inhomo13}(top plots), it is clear that when the density is inhomogeneous inside a body, with constant density outside $\rho_{G}$,
the sharp gradient of the scalar field near the surface is smaller than considering constant density inside.
Also when the density is inhomogeneous inside and connected smoothly to $\rho_{G}$ outside,($\rho_{s}$ is the surface density  
and $\rho_{c}$,$\rho_{0}$ and $\lambda$ are some constants) 
the field gradient is much smaller than two previous cases and the scalar field is closer to it's minimum in these layers.
This effect is getting smaller by choosing smoother density profiles figure~\ref{fig:inhomo13}(bottom plots).
Therefore, in a nonhomogeneous sphere with smooth density change, the thin shell features mentioned above 
will be weaken. 
This effect might be important in bodies with extended atmosphere where the density is smooth with very small $\frac{d\rho}{dr}$ like Red Giant
 stars.

To draw a conclusion, we found that the above two features of thin unscreened layer are artifact of the ideal model of constant density 
inside sphere (homogeneity) where induces large density variation at the surface.
 we can deduce that by following the Khoury and Weltman approach in static structures,
 the field profile in inhomogeneous density distribution which is somehow a more realistic case is 
different from homogeneous distribution in a way that the field is closer to its minimum as density decreases towards surface of the structure.
 We should note that this does not mean that the whole star might be screened. Variable density will induce variable $\phi_{min}(r)$ 
which imprints a gradient on $\phi(r)$ between the layers and makes the whole star unscreened. But we have shown that
this fifth force is  weaker than what peoples deduce from homogeneous assumption in the thin shell near the surface.

\section{Realistic model}
In order to study the stellar evolution we have used MESA code. Modules for Experiments in Stellar Astrophysics, MESA is a 1D stellar evolution 
module for simulation of stars from pre main sequence to the final stages. MESA solves the fully coupled stellar structure equations \cite{32}. 

Since the equations in MESA are all arranged in cgs units, it is necessary to convert field equations from natural units to cgs.
The starting point is the action, which has the dimension $ML^2T^{-1}$.

\[  S[ML^2/T]= \int d^{4}x \sqrt{-g} \bigg( \frac{M_{P}^{2}}{2}R - \frac{1}{2}\partial_{\mu}\tilde{\phi}\partial_{\nu}\tilde{\phi} g^{\mu\nu}-V(\tilde{\phi}) \bigg),   \]
where the metric is dimensionless, $R$, $d^{4}x$ and $\partial_{\mu}$ have the dimensions $L^{-2}$, $L^4$ and $L^{-1}$ respectively. Thus
 $M_{P}^{2}$ must turn to $M_{P}^{2}=\frac{c^3}{8 \pi G}$ with the dimension $\frac{M}{T}$.
The scalar field dimension can be found by the kinetic term $K= \int d^{4}x \sqrt{-g}(- \frac{1}{2}\partial_{\mu}\tilde{\phi}\partial_{\nu}\tilde{\phi} g^{\mu\nu})$
 in the action
\[ K[ML^2/T]= \frac{L^{4} \tilde{\phi}^{2}}{L^{2}}.  \]
Therefore, $\tilde{\phi}$ has the dimension $M^{\frac{1}{2}}T^{\frac{-1}{2}}$ as $M_{P}$ or we can make it dimensionless by introducing a new scalar field $\phi
= \frac{\tilde{\phi}}{ M_{P}} $. Returning to the Lagrangian, the scalar potential $V(\tilde{\phi})$ must have the dimension $ML^{-2}T^{-1}$.

\subsection{Chameleon Scalar field with Ratra-Peebles potential}

The scalar-tensor action for the chameleon  field in the Einstein frame is given by 
\begin{equation}
S= \int d^4x \sqrt{-g} \bigg( \frac{M_P^2}{2}R - \frac{M_P^2}{2}\partial_\mu \phi\partial_\nu\phi g^{\mu\nu}-V(\phi) \bigg) + S_m(g_{\mu\nu}A^2(\phi),\psi_{i}),
\end{equation}
where $A(\phi)$ is coupling function.
The scalar field equation of motion can be found by the variation of the Lagrangian with respect to the field $\phi$. Therefore we have
\begin{equation}
M_{P}^{2}\square\phi = \frac{\partial V}{\partial \phi}- \frac{T_{m}}{c}\frac{A_{,\phi}}{A},
\end{equation}
where the self-interaction potential is of the form $V(\phi)=M^{4+n }\phi^{-n} M_{P}^{-n}$ and 
the coupling function $A(\phi)=e^{\beta\phi}$, with $\beta=\frac{A{_{,\phi}}}{A}$, which can be approximated by the linear form
 $A(\phi)\simeq 1+\beta\phi$ where $\beta$ denotes the coupling to
 matter. Assuming the major component of the matter is non relativistic  then $T_{m}=-\rho c^2$, where $\rho$ is 
the conserved mass density in the Einstein frame and the field equation of motion turns to be 
\begin{equation}
\square\phi = \frac{1}{M_{P}^{2}}\frac{\partial V}{\partial \phi}+\frac{\beta\rho c}{M_{P}^{2}},
\end{equation}

Comparing This equation of motion with standard massive Klein Gordon equation of motion we can introduce an effective potential 
\begin{equation}
V_{eff} = \frac{1}{M_{P}^{2}}V(\phi) + \frac{\rho c}{M_{P}^{2}}(1+\beta \phi).
\end{equation}
This effective potential has an absolute minimum that is related to the local density of matter which we denote it by $\phi_{min}$
\begin{equation}
\phi_{min}= \bigg(\frac{nM^{4+n}}{M_{P}^{n}\rho \beta c }  \bigg)^{\frac{1}{1+n}}.
\end{equation}
small oscillations around this minimum induce a mass to the scalar field which is the key factor in field interaction with the environment to reproduce screening behavior.
 The squared mass can be calculated from the value of second derivative of effective potential at its minimum which is of the form
\begin{equation}
m_{eff}^{2}=\frac{\partial^{2} V_{eff}}{\partial \phi^{2}}\lvert_{\phi_{min}} = \frac{n(n+1) M^{4+n}}{M_{P}^{n+2} \phi_{min}^{n+2}} .
\end{equation}
In addition, it is possible to constrain the constant $M$ by requiring that the transition from deceleration to acceleration of the expanding  universe 
must be consistent with our $\Lambda$ dominated era \cite{30}. Although tests of general relativity in
the solar system would exclude the corresponding parameter space for the chameleon field 
 but in order to find significant deviations for the dynamical chameleon scalar field
 with attractor behaviour
\footnote{By Attractor behaviour we mean the scalar field which follows the minimum of the effective potential in each layer in
 inhomogeneous density distribution since in Section3.3 it is shown that this scalar field is fast enough to follow the minimum during RGB evolution 
timescale.}
in astrophysical scales we have focused on greater values of the constant taken from cosmology.
It's order of magnitude for different values of $n$ can be found using the following relation \cite{33,34}
\begin{equation}
\label{eq:const}
logM[GEV]\approx\frac{19n-47}{4+n}
\end{equation}

Also, there are two cosmological attractor solutions for the quintessence scalar field with inverse power law potential to follow the behavior 
of the dominant constitute of the universe: a tracking solution  (with $\omega_{\phi}=-1+\frac{n}{n+2}$ , $ \Omega_{\phi}<<1$) in radiation 
dominated and matter dominated eras and a dark energy dominated attractor  (with $\omega_{\phi}\rightarrow-1$ , $ \Omega_{\phi}=1$).
 A numerical analysis shows that the bound $\omega_{\phi}<-0.7$ today, can be
 satisfied for $ n \leq 2$ \cite{31}.

\subsection{ Hydrostatic Equilibrium equation in scalar-tensor theories}

The effect of fifth force in modified gravity theories can be added to the Hydro Static Equilibrium (HSE) equation of the star by introducing
 the scalar field radial acceleration $g_{\phi}$. The hydrostatic equilibrium equation of a star is given by

\begin{equation}
\frac{-1}{\rho}\frac{dP}{dr}+ g_{eff}=0, 
\end{equation}
where $g_{eff}$ is combination of scalar and Newtonian radial accelerations ($g_{eff}=g_{\phi}+g_{N}$).  In order to find $g_{\phi}$ 
in cgs units, we implement  conservation of energy-momentum tensor.  In conformal scalar-tensor theories, the energy-momentum tensor is
 not conserved in the Einstein frame which is given by
\begin{equation}
\nabla_{\mu}T^{\mu\nu}=\beta(\phi)T_{m}\nabla^{\nu}\phi.
\end{equation}

 In weak field limit, the conservation equation is reduced to nonrelativistic geodesic equation
\begin{equation}
(\rho) (\ddot{x}^{a}+\Gamma^{a}_{bc}\dot{x}^{b}\dot{x}^{c})= \beta(\phi)(-\rho c^2)\nabla^{a}\phi,
\end{equation}
and finally revealed the form of extra fifth force in Newton second law  
\begin{equation}
\frac{d^{2}x^{i}}{dt^{2}}= -\nabla\Phi-\beta c^2 \nabla\phi.
\end{equation}
where $g_{N}=-\nabla\Phi$ and $\Phi$ is the Newtonian gravitational potential and the last term on the right side of the equation is
 considered as the scalar field radial
acceleration $g_{\phi}$. Therefore, the HSE equation in the Eulerian framework is
\begin{equation}
\label{eq:HSE1}
\frac{dP}{dr}=- \rho \bigg(\frac{G M_{r}}{r^2}+\beta c^2 \nabla\phi \bigg),
\end{equation}
and in the Lagrangian framework
\begin{equation}
\label{eq:HSE2}
\frac{dP}{dm}=-  \bigg(\frac{G M_{r}}{4 \pi r^4}+\rho\beta c^2 \frac{d\phi}{dm} \bigg).
\end{equation}
where both sides have the dimension of  $L^{-1}T^{-2}$ in cgs units.

\subsection{Attractor Solution}
When a normal star leaves its main sequence stage its unstable core contracts and releases a huge amount of gravitational and nuclear energy.
 This energy expands the outer atmosphere of the star and makes a red giant. In our modified gravity theory this evolution must shift the minimum
 of the effective potential to the higher values of the scalar field. If we assume that the field sits in its potential minimum in the main sequence
 stage the question is "how fast is the scalar field to roll down to the new potential minimum?" or "Is it reasonable to assume that in evolution
 from main sequence to giant the scalar field is always relaxed in its potential minimum since the rolling down  occurs in a faster rate with respect
 to the stellar evolution?"

The response time $t_\phi$, for the chameleon is determined by the period of oscillation about the minimum which is equal to
 $(m_{eff} c)^{-1}$ \cite{29}.
Time scale of stellar evolution is given by free fall time $t_{ast}$, where for a star with average density $\rho$ it could
 estimated to be $t_{ast} = (G\rho)^{-\frac{1}{2}}$. 
In this paper, we focus on the  RGB phase and it is assumed that the scalar field has relaxed to the minimum of the effective potential
 during the long period of main sequence phase. To explore the effect of modified gravity, the HSE equation of MESA code has been changed 
according to section 3.2. To estimate $t_{ast}$ we used MESA dynamical timescale(the smallest timescale in stellar evolution);
 for the scalar field to follow the minimum, 
the attractor condition $t_{\phi}<< t_{ast}$ must be satisfied.

\begin{figure}[tbp]
\centering
\includegraphics[scale=0.50]{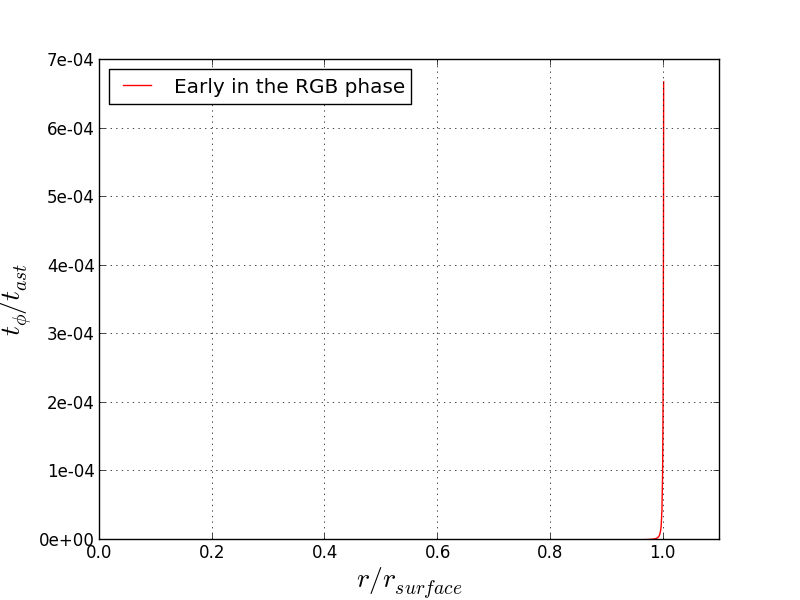}
\includegraphics[scale=0.50]{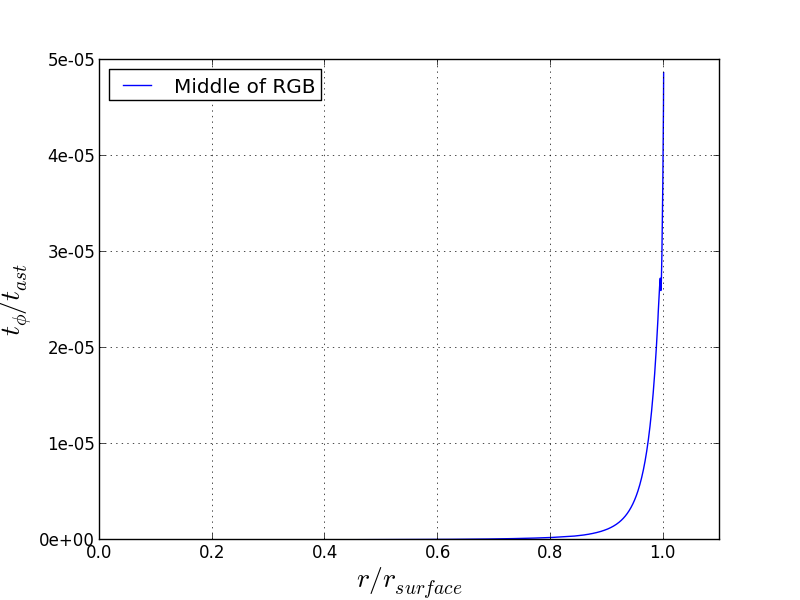}
\includegraphics[scale=0.50]{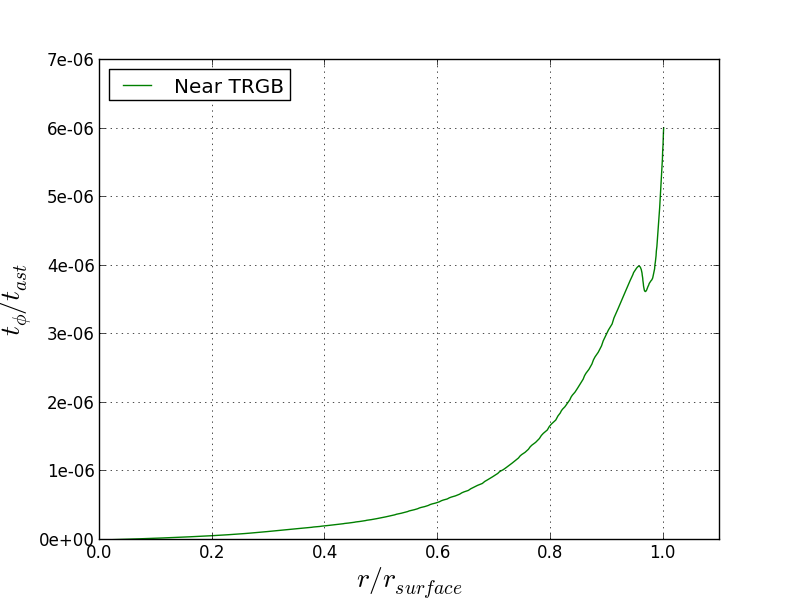}
\caption{\label{fig:2}The timescale for the scalar field oscillation about the minimum $t_{\phi}$ over the timescale 
for the stellar evolution in RGB phase
 $t_{ast}$ is plotted as a function of star's radius for three times during the stellar evolution .} 
\end{figure}

To gain some insight into the order of $t_{\phi}$ and $t_{ast}$ in different stages of stellar evolution, 
we have simulated a one solar mass star with metallicity $Z=0.02$, mixing parameter $\alpha=2$, Reimers and Blockers
 mass loss parameters $\eta_{R}=0.5$,  $\eta_{B}=0.1$,
  which is governed by a chameleon modified gravity model with $n=1$,
 $ M\simeq2.5118\times 10^{-6} (GeV) $ and
 $ \beta=1/ \sqrt6 $ from pre main sequence phase until Tip of the Red Giant Branch(TRGB).
 We have explored three different epochs in the stellar evolution to investigate the attractor condition: i) early in the RGB phase, ii) in the middle of the RGB evolutionary track and iii) at the TRGB.
The ratio between chameleon evolution and star evolution  time scales  $(\frac{t_{\phi}}{t_{ast}}) $ as a function of star's
 radius for three mentioned epochs is shown in figure~\ref{fig:2}.

\begin{figure}[tbp]
\includegraphics[scale=0.75]{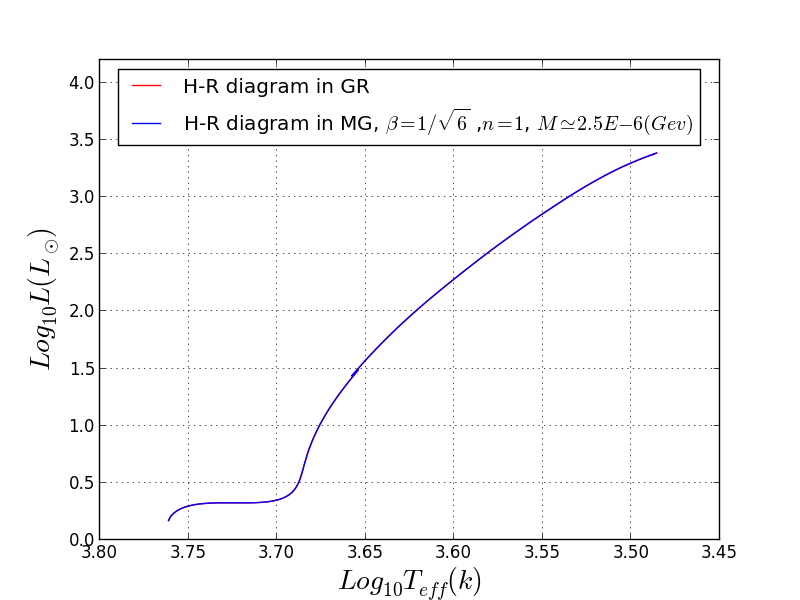}
\includegraphics[scale=0.75]{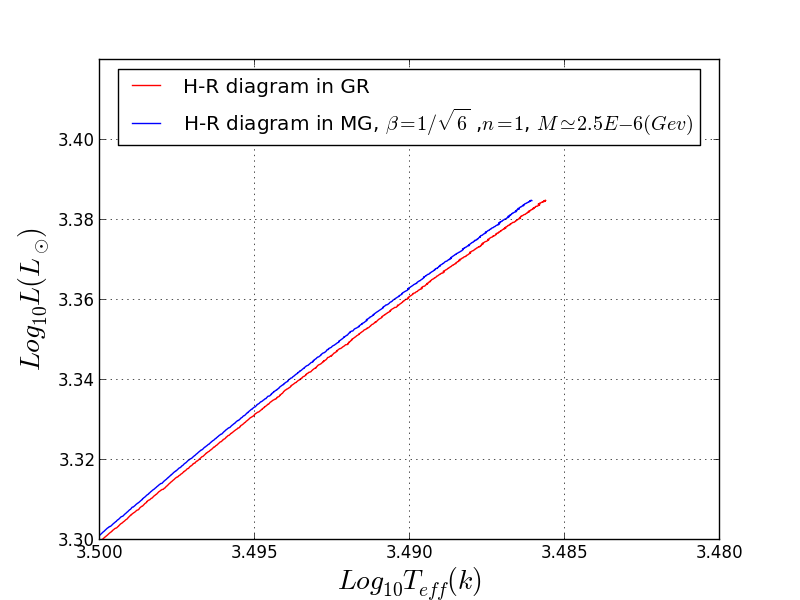}
\caption{\label{fig:3} The top plot shows the evolutionary track of one Sun mass star in the Red Giant Branch of the HR diagram. The bottom plot indicates the Tip of the Red Giant Branch.}\label{HRdiag1}
\end{figure}

It is obvious from the plots that $\frac{t_{\phi}}{t_{ast}}$ is less than unity in the RGB phase.
It implies that the oscillation of the scalar field around the effective potential is fast enough compared to the stellar evolution, to consider 
the scalar field in the minimum of the effective potential on average. Therefore, it is legitimate to impose the condition that the field is 
relaxing in the minimum during whole RGB phase. By the way it would be easy to calculate the fifth force inside the red giant. The extra force 
term in hydrostatic equilibrium Eqs \eqref{eq:HSE1}, \eqref{eq:HSE2} can be calculated from gradient of density inside  the star. 
To see how this fifth force, extracted from an attractor solution can affect the red giant phase of stellar evolution, we modified HSE equation  
inside the MESA and run it for a one solar mass star to see the difference in evolution in the presence of dynamical chameleon field
\footnote{Although to study the behaviour more accurately, one should solve the scalar field full time- and space- dependent
equation of motion inside MESA code}.

figure~\ref{fig:3} shows the result. It seems that modified gravity effect is negligible during the red giant phase due to weak fifth force
 probably result in from small gradient in density inside the RGB considering dynamical chameleon scalar field.
 This is in contradiction with the quasi-static approximation considered 
in paper \cite{21}, Chang and Hui. Where there is an unscreened mantle in the exterior envelope of the RGB star and the effect of fifth force 
is noticeable there. Therefore comparing our result for $\beta=\frac{1}{\sqrt{6}}$ with figure~3 of the paper \cite{21},
 insignificant deviations appear in the evolutionary track of RGB star in the
 HR diagram from the GR one for chameleon scalar field with attractor behaviour.

\section{Constraining Chameleon Model with Attractor behaviour}

\begin{figure}[tbp]
\centerline{\includegraphics[width=15cm,height=70cm,keepaspectratio]{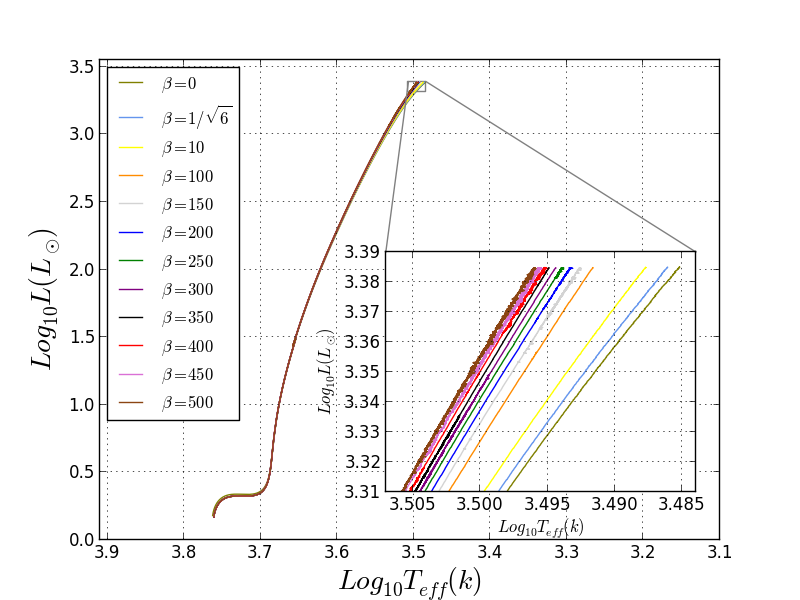}}
\caption{\label{fig:4} the HR diagram for different coupling constants with a closer look at Tip of the Red Giant Branch.
The instability in the evolutionary track for $\beta=400$(red line) is clear which is intensified for $\beta=450$(purple line) 
and $\beta=500$(brown line).}
\end{figure}

\begin{figure}[tbp]
\includegraphics[scale=0.70]{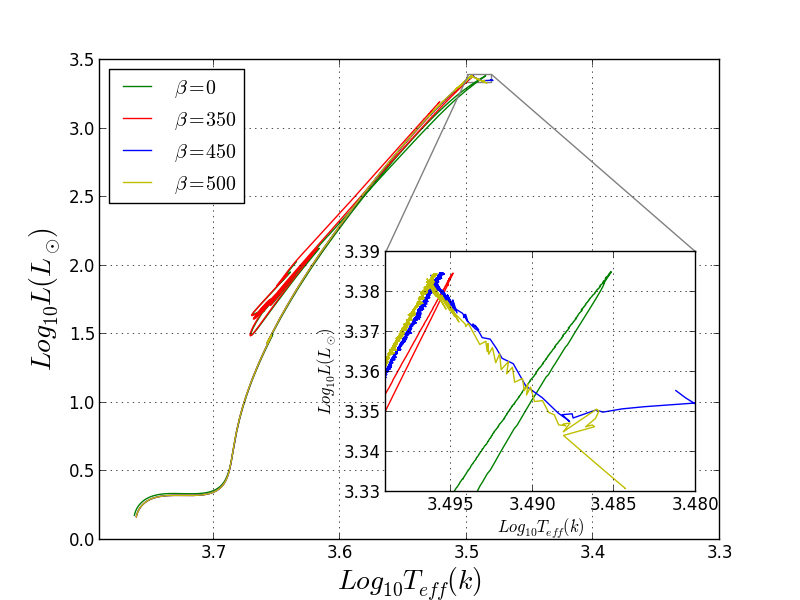}
\includegraphics[scale=0.70]{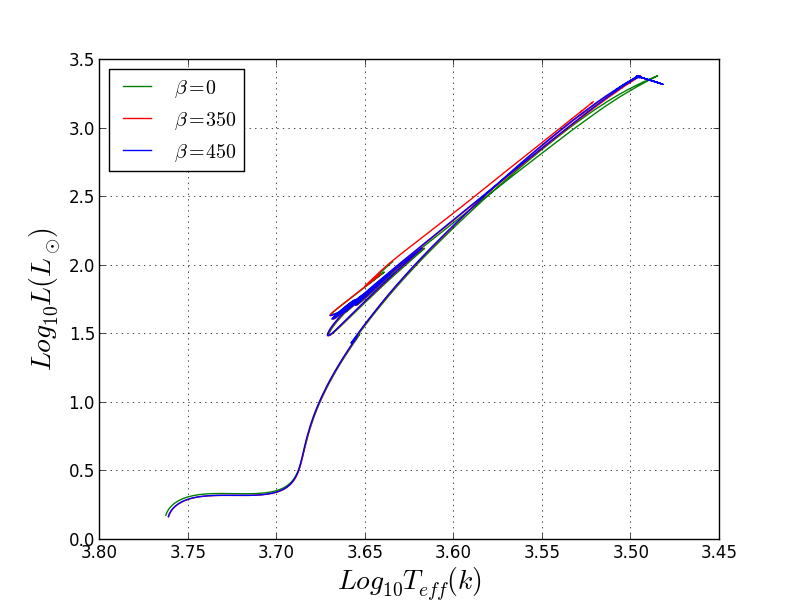}
\caption{\label{fig:5} Top plot: HR diagram for $\beta=0$,$\beta=350$,$\beta=450$,$\beta=500$. Bottom plot: The Red Giant Branch for $\beta=450$
 with increased spatial resolution.}
\end{figure}

In this section, we will study the stability of nonrelativistic stars in the presence of chameleon 
scalar field relaxed on it's potential minimum, as a test of modified gravity and also try to constrain the coupling constant $\beta$ by tracing 
evolution of a one solar mass RGB star using MESA. To do so, we increase $\beta$ in the code and check if it can make the star unstable or the 
evolution of the star deviates significantly from the typical GR track and if the TRGB temperature changes to the values which are out of the 
observed range. figure~\ref{fig:4} shows the HR diagram for different coupling constants. It is obvious that by increasing $\beta$, the evolutionary 
track deviates more from the typical GR  track ($\beta=0$) and for about $\beta\gtrsim150$ it seems 
that there are some instabilities inside the star. As $\beta$ increases, instabilities in the evolutionary track increase in such a way that 
the star leaves the standard 
track and finally the code fails to converge and terminates for $\beta = 450,500$, figure~\ref{fig:5} (top plot).

\begin{figure}[tbp]
    \centering
\includegraphics[width=0.45\textwidth]{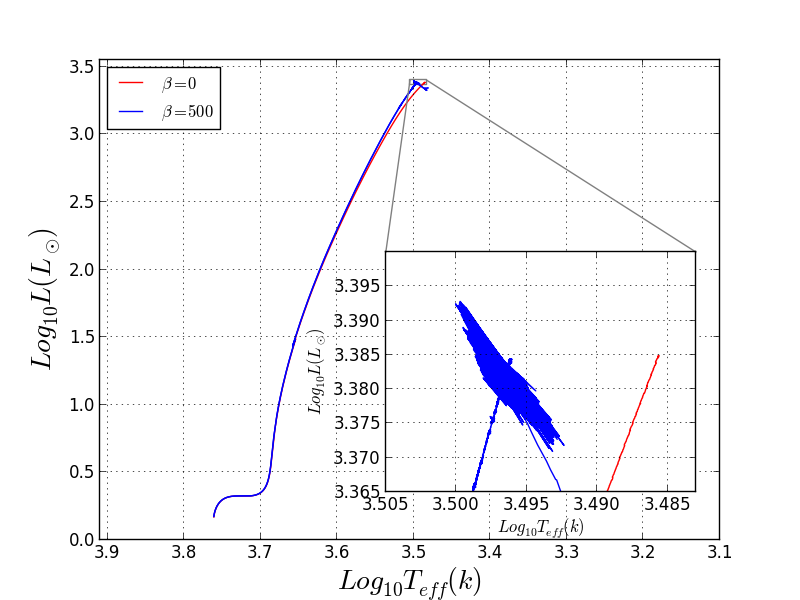}
\hfill
     \includegraphics[width=0.45\textwidth]{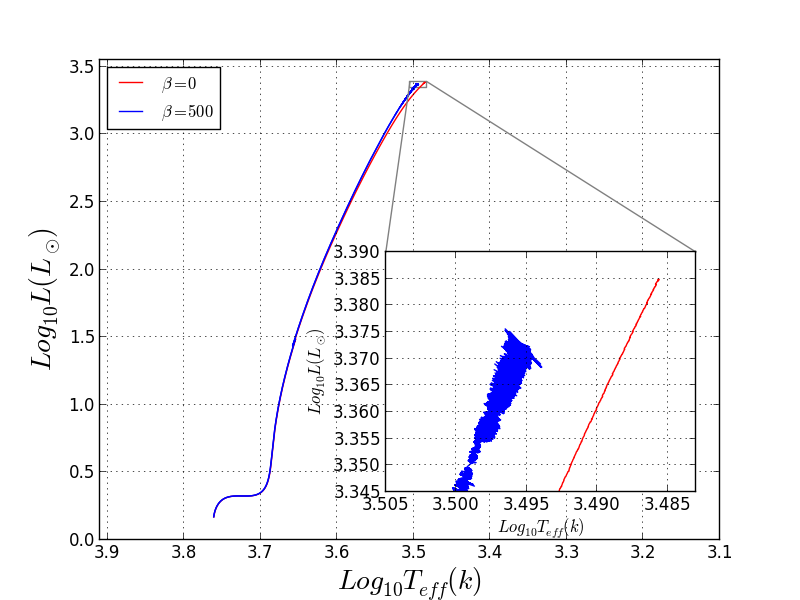}
\hfill     
    \includegraphics[width=0.45\textwidth]{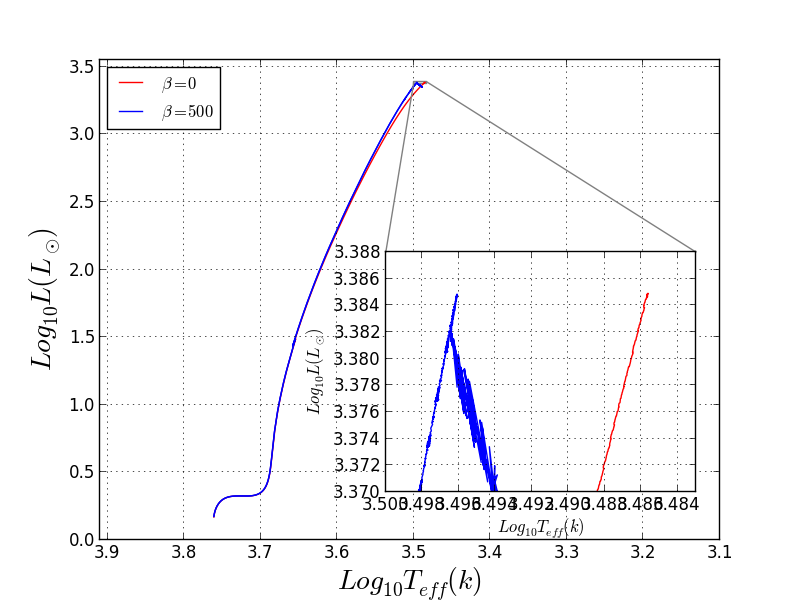}
 \hfill   
 \includegraphics[width=0.45\textwidth]{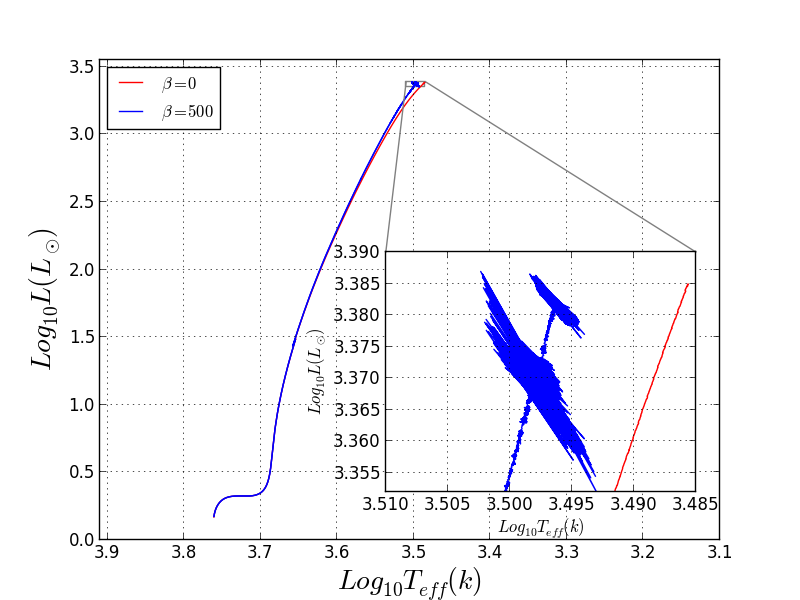}
 \hfill
 \includegraphics[width=0.45\textwidth]{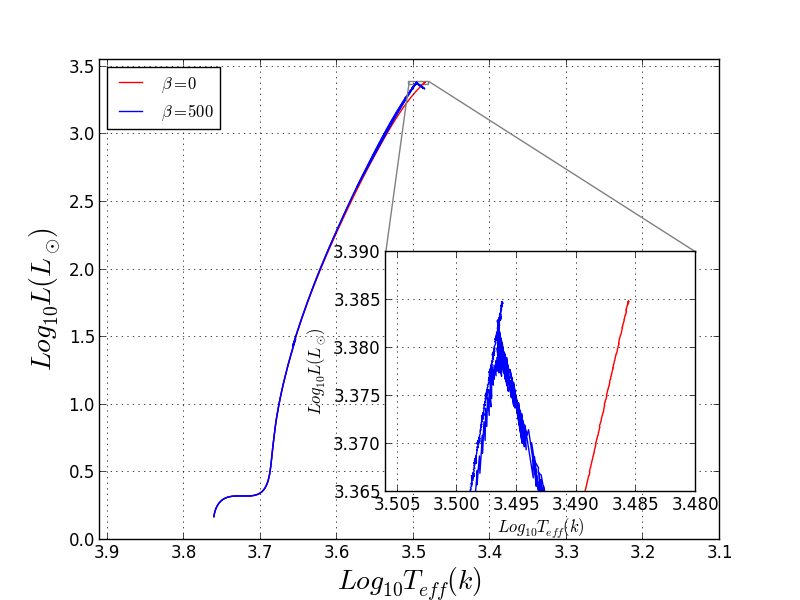}

     \caption{\label{fig:6}HR diagram for $\beta=500$ with different resolutions. In the first row,
 $ varcontrol-target=10^{-4},  mesh-delta-coeff=0.5$ (left), 
 $ varcontrol-target=10^{-4},  mesh-delta-coeff=0.009$ (right),
in the second row, 
$ varcontrol-target=10^{-3},  mesh-delta-coeff=0.5$ (left), 
$ varcontrol-target=10^{-2}, mesh-delta-coeff=1$ (right),
and in the third row, $ varcontrol-target=10^{-1}, mesh-delta-coeff=0.7$.
 As it is clear, Although the instability in the evolutionary 
    track is decreased in some resolutions, it is not totally resolved. }
  \end{figure}

\begin{figure}[tbp]
    \centering
    \includegraphics[width=0.6\textwidth]{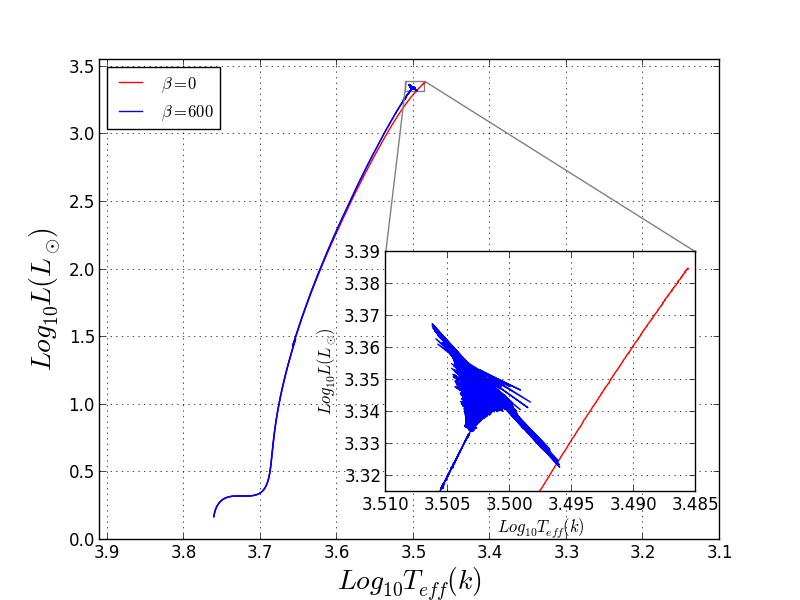}
    \includegraphics[width=0.6\textwidth]{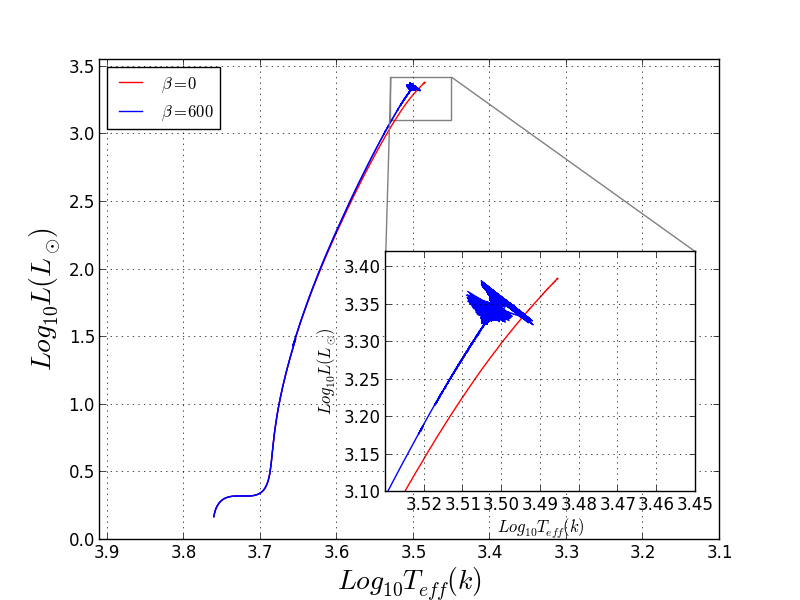}
 \hfill   
    \includegraphics[width=0.6\textwidth]{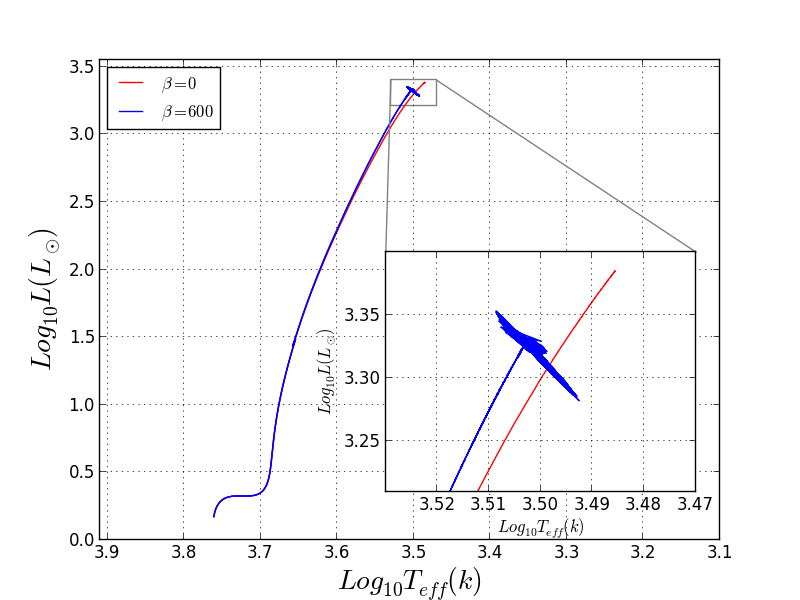}
    \hfill
    \caption{\label{fig:7}HR diagram for $\beta=600$ with different resolutions.
   In the first row, $ varcontrol-target=10^{-3}, mesh-delta-coeff=0.5$, In the second row, $ varcontrol-target=10^{-2},  mesh-delta-coeff=1$.
   In the third row, $ varcontrol-target=10^{-1}, mesh-delta-coeff=0.7$.
 The instability in each resolution is increased compared 
 to the same resolution for $\beta=500$. }

    \end{figure}

The question then arises: whether this is a real, physical instability in the evolutionary track of the star or it is induced by numerical instability in the code.
In order to answer this question we changed the spatial and temporal resolution in MESA code for $\beta=450$ and $\beta=500$ by changing 
"mesh-delta-coeff" and "varcontrol-target". These parameters adjust the resolution according to maximum allowed changes in structural quantities 
from cell to cell or in a time step in MESA code.
The results in figure~\ref{fig:5} (Bottom plot) shows that the instability at the end of RGB phase for $\beta=450$ decreases by changing  
the resolution (not totally removed) and the star continues it's evolution on the horizontal branch.
Therefore, it seems that by changing resolution, it might be possible to resolve the instability and find out the actual effect of 
fifth force on the stellar evolution.
So, next step was running MESA  and finding the resolution for which larger $\beta s$ could converge with less numerical instability.
It was done for $\beta=500, 600 ,700, 800, 900, 1000$. We have chosen varcontrol-target = $10^{-4}$, $10^{-3}$, $10^{-2}$, $10^{-1}$ with
mesh-delta-coeff = 1,0.5,1,0.7 and also 
varcontrol-target = $10^{-4}$, mesh-delta-coeff = 0.5, 0.009. For $\beta=500$ we found that still there are some instabilities in the evolutionary track which are shown in figure~\ref{fig:6} for some resolutions.
Moreover, $\beta=600$ converged for the resolutions mentioned in figure~\ref{fig:7}, it is clear that changing the 
resolution does not eliminate the instability.

\begin{figure}[tbp]
\centerline{\includegraphics[scale=0.70]{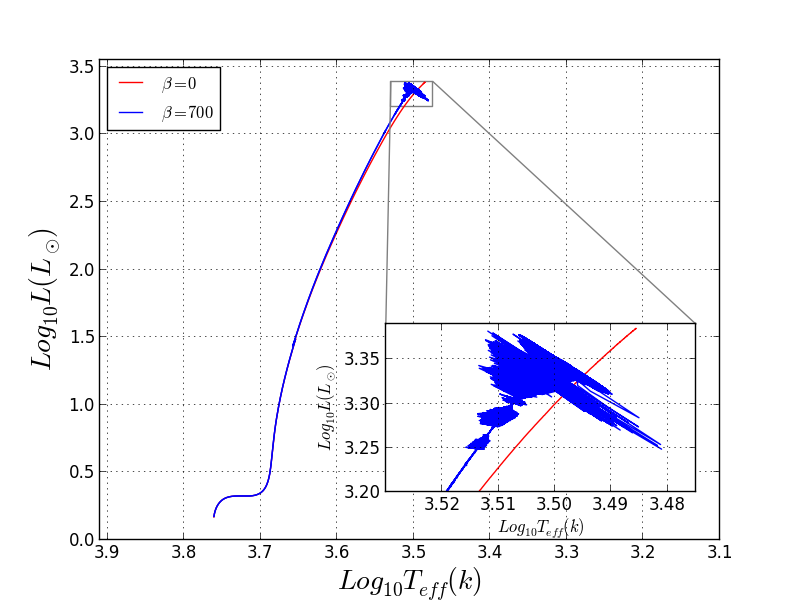}}
\centerline{\includegraphics[scale=0.70]{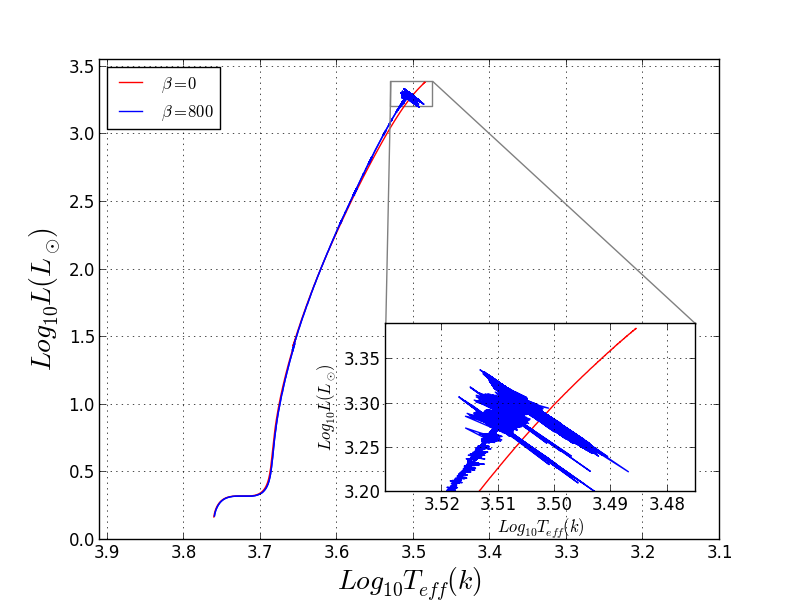}}
\caption{ \label{fig:8}Top plot: the HR diagram for $\beta=700$ with $ varcontrol-target=10^{-2}$ , $  mesh-delta-coeff=1$, Bottom plot: the 
HR diagram for $\beta=800$ with 
$ varcontrol-target=10^{-3}$ \newline $  mesh-delta-coeff=2.3$.}
\end{figure}

\begin{figure}[tbp]
\centerline{\includegraphics[scale=0.70]{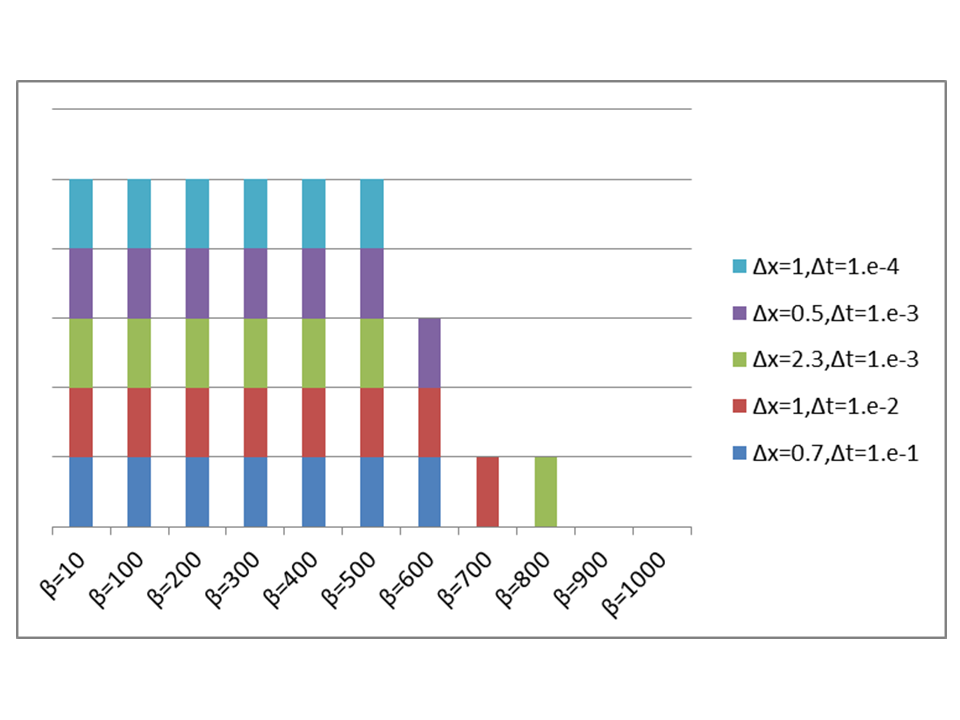}}
\caption{\label{fig:9}Plot shows the convergence resolutions of the code for different coupling constants. $\Delta t$ and $\Delta x$ stand for $"varcontrol-target"$ and $"mesh-delta-coeff"$ respectively. It is clear that by increasing $\beta$,
the code converges for fewer numbers of resolution which indeed demonstrates the increased instability.}
\end{figure}

Also, in the most stable form of the code for $\beta=700,800$, we found just one resolution for convergence of the code, figure~\ref{fig:8}.
 By comparing the plots for different coupling constants it can be understood that by increasing $\beta$ indeed the instability increases in the plots.
It should be noted that $\beta<400$ converged for all the resolutions mentioned above.
 Moreover, no resolution found that makes bigger values of $\beta$(>800) converging from pre main sequence. 
 A schematic view of the convergence range of the code for 
different choices of resolution and coupling constant is represented in figure~\ref{fig:9}.
It can be understood that, the convergence range is getting more limited by increasing $\beta$ which is 
an evidence for the growth of instability.

\begin{figure}[tbp]
\centerline{\includegraphics[scale=0.70]{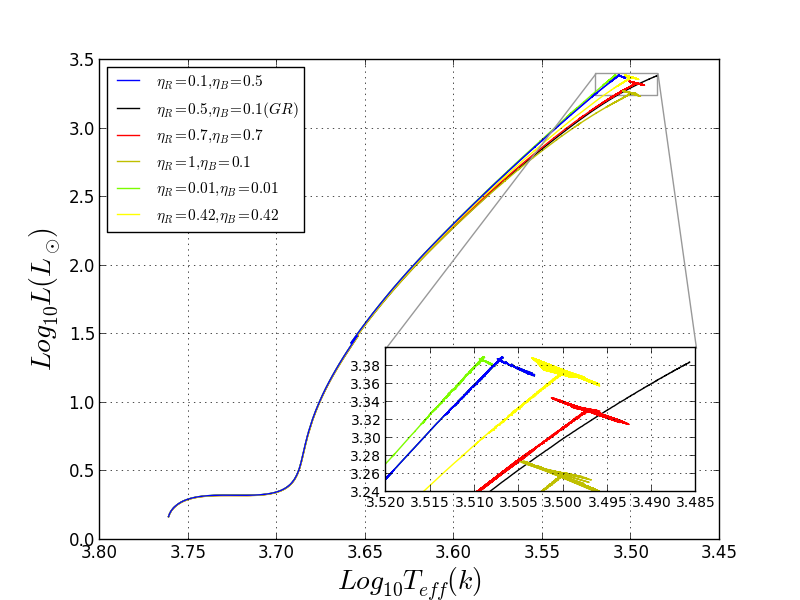}}
\centerline{\includegraphics[scale=0.70]{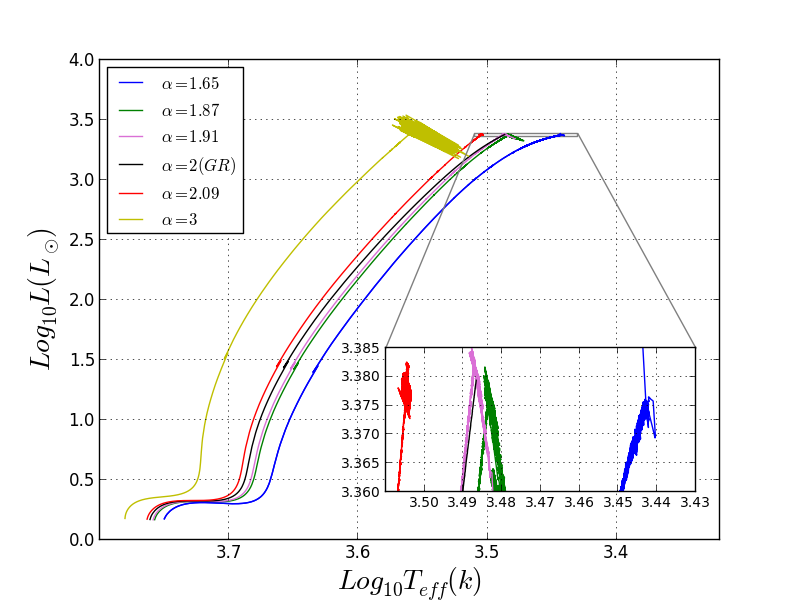}}
\caption{\label{fig:10} The top plot shows the evolutionary track of one solar mass star with $\beta=500$ in the Red Giant Branch of the HR diagram for
 different values of Reimers and Blockers mass loss parameters $\eta_{R}$, $\eta_{B}$ respectively.
 The bottom plot indicates the evolutionary track for different values of mixing parameter $\alpha$.}
\end{figure}

In addition, in order to check how the results are sensitive to mass loss and
mixing length we have changed these parameters in MESA code for $\beta=500$. According to
figure~\ref{fig:10} changing these parameters cannot remove the instability.

\begin{figure}[tbp]
    \centering
\includegraphics[width=0.70\textwidth]{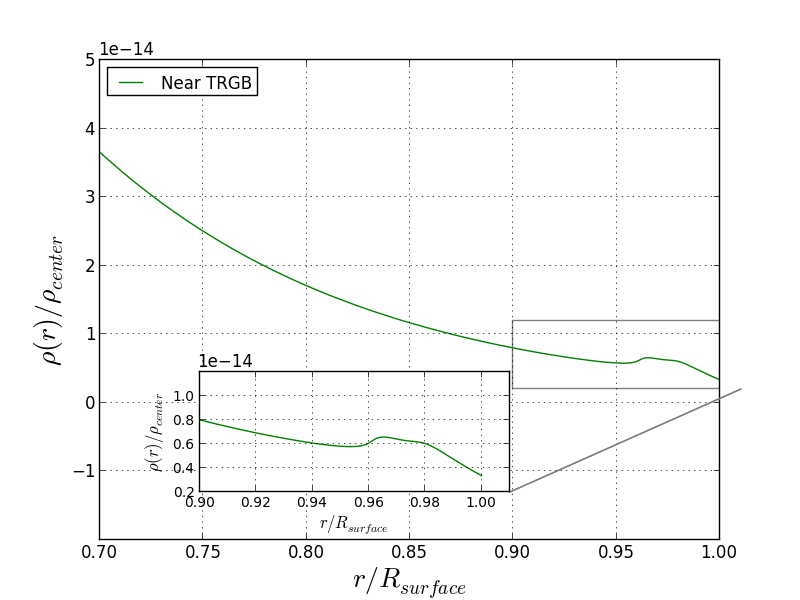}
\hfill
     \includegraphics[width=0.70\textwidth]{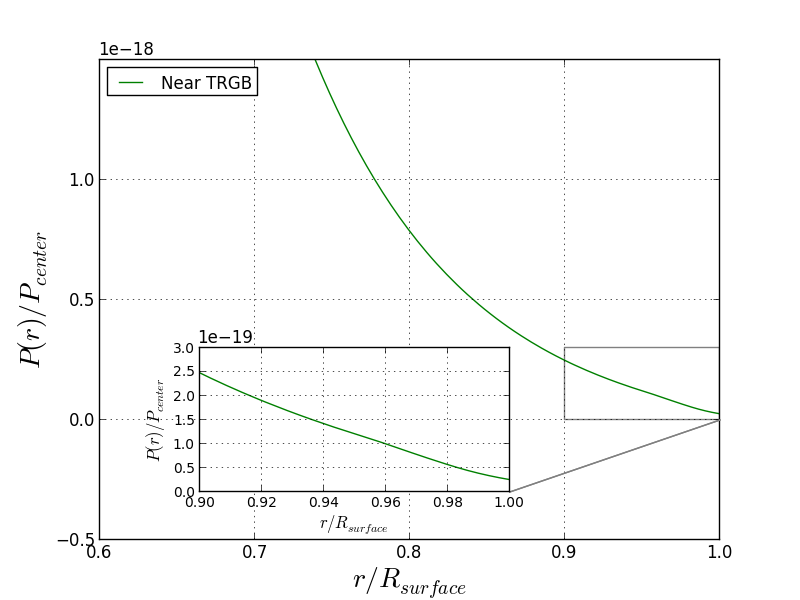}

 \caption{\label{fig:grbump}The density and pressure of the star are plotted as a function of radius near TRGB in GR model.}
  \end{figure}

It might be reasonable to explain the instability according to RGB density distribution.
The existence of density inversion in the
outer envelope of RGB stars is discussed in \cite{36} in detail.
If we look at MESA density distributions in GR case it is obvious that there are density bumps
in RGB phase approximately from the middle of RGB track to the top. The density and pressure
in one snapshot of star near TRGB are indicated in figure~\ref{fig:grbump} in GR model. As it is clear
 density inversion arose near the surface to preserve Hydrostatic Equilibrium therefore no pressure inversion occurs in GR case.

In our model, Since the fifth force is restricted to the gradient between the minimums, it is directly related to the density gradient.
 If there were no density bump, the modified gravity force was 
attractive in all layers. In the place of density bump, the fifth force changes sign and becomes 
repulsive. In fact, density bump 
induces a similar and inverted bump on the field.
The fifth force is attractive
when the field is rising and repulsive when it is setting.

In most of the papers that assumed quasi-static approximation,
the scalar field equation of motion is approximated as


\begin{equation}
\nabla^{2}\phi = \frac{\beta\rho c}{M_{P}^{2}}.
\end{equation}
Therefore the total
force per unit mass (gravity and modification) in the unscreened region would be

\begin{equation}
F = \frac{-GM}{r^{2}}\Bigg(1+2\beta^{2}\Bigg(1-\frac{M(r_{s})}{M(r)}\Bigg)\Bigg)
\end{equation}

And the effect of modified gravity is then changed to an 
effective gravitational constant $G \rightarrow G(r) \equiv  G(1 + 2\beta_{eff}^{2})$ in the unscreened region.
As it is clear, the modified gravity force would always be attractive using this approximation.
 If the parameters are chosen in such a way that the static chameleon 
scalar field follows the potential minimum in each layer, by considering this approximation, the density bump would never cause repulsive force.

This repulsive force is important for us since the chameleon scalar field with attractor behavior follows it's minimum in each layer.
The repulsive behaviour
could be destructive to the stellar structure and make real physical instability.

\begin{figure}[tbp]
    \centering
\includegraphics[width=0.45\textwidth]{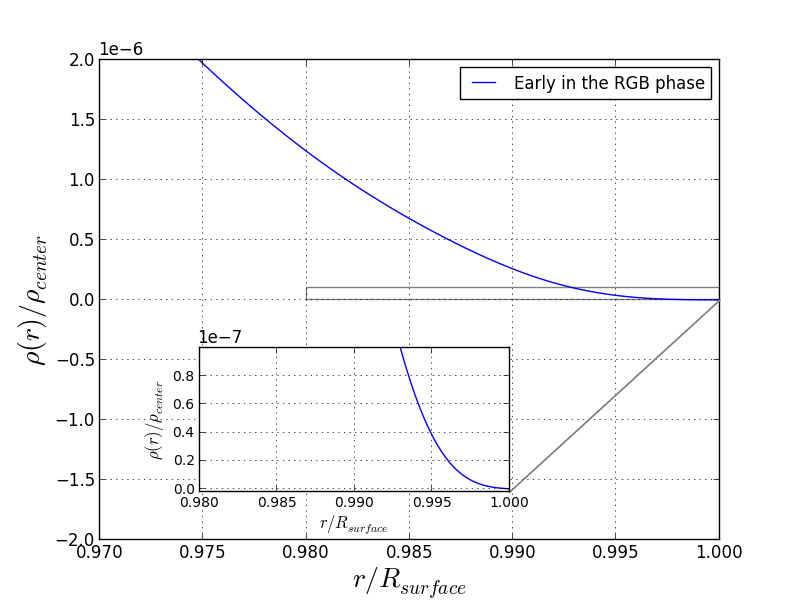}
\hfill
     \includegraphics[width=0.45\textwidth]{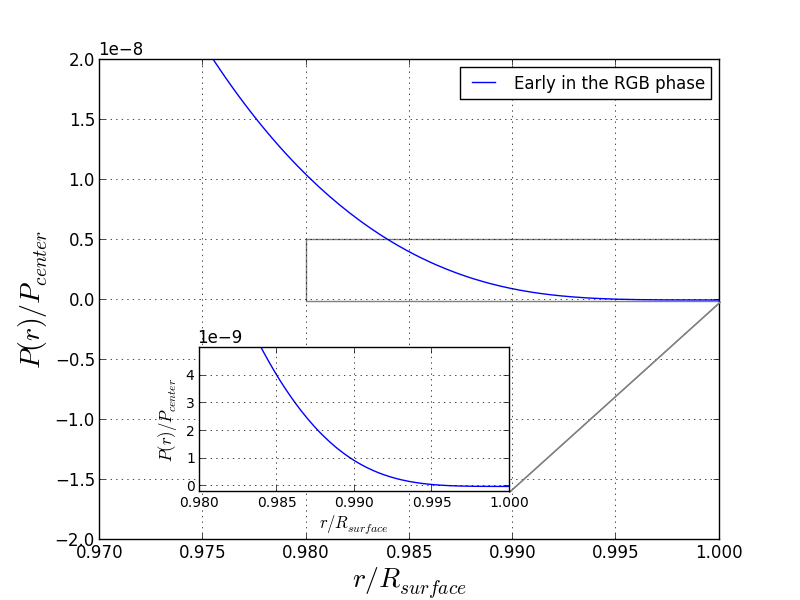}
\hfill     
    \includegraphics[width=0.45\textwidth]{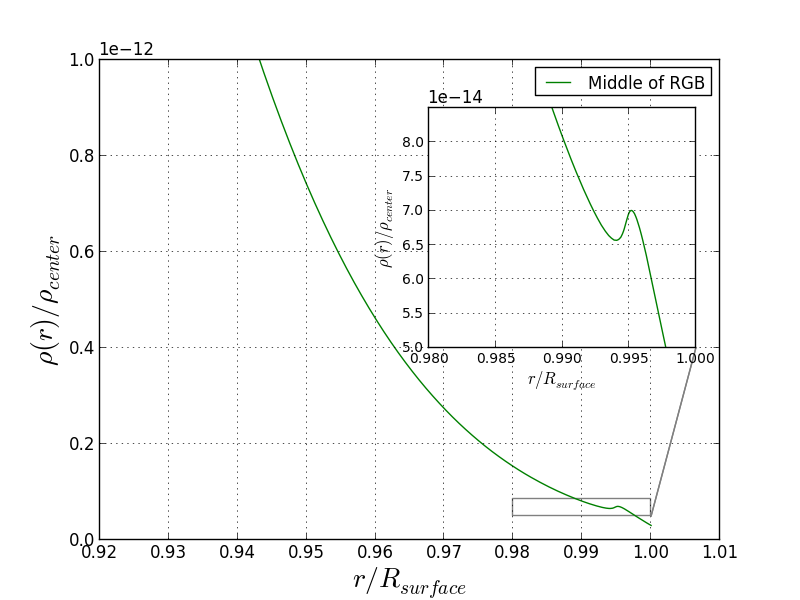}
 \hfill   
 \includegraphics[width=0.45\textwidth]{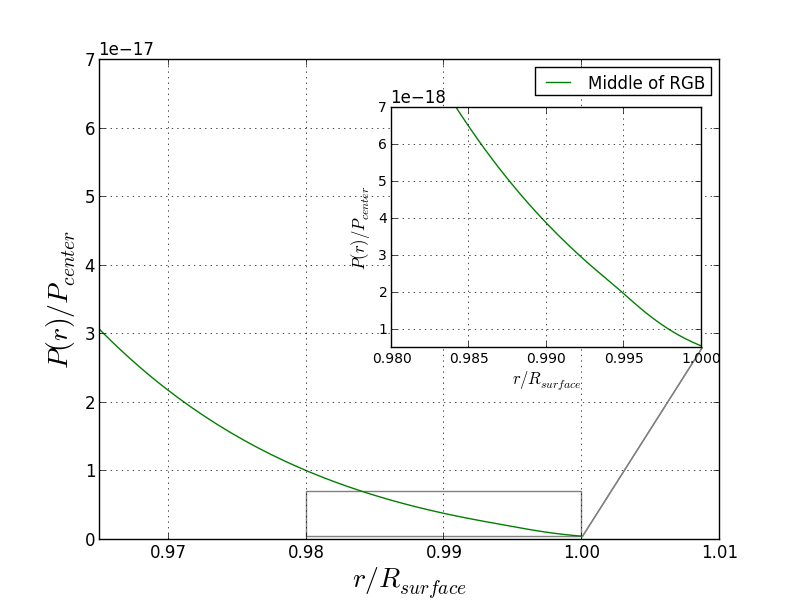}
 \hfill
 \includegraphics[width=0.45\textwidth]{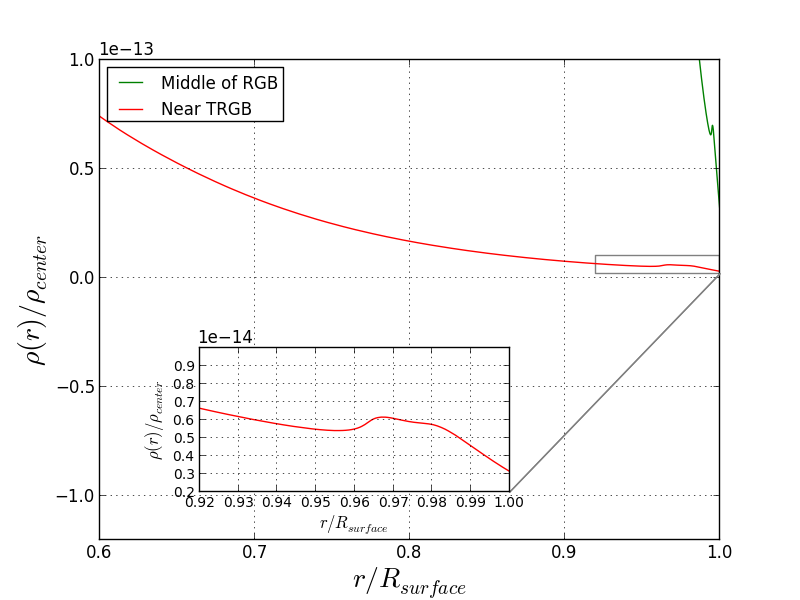}
  \hfill
 \includegraphics[width=0.45\textwidth]{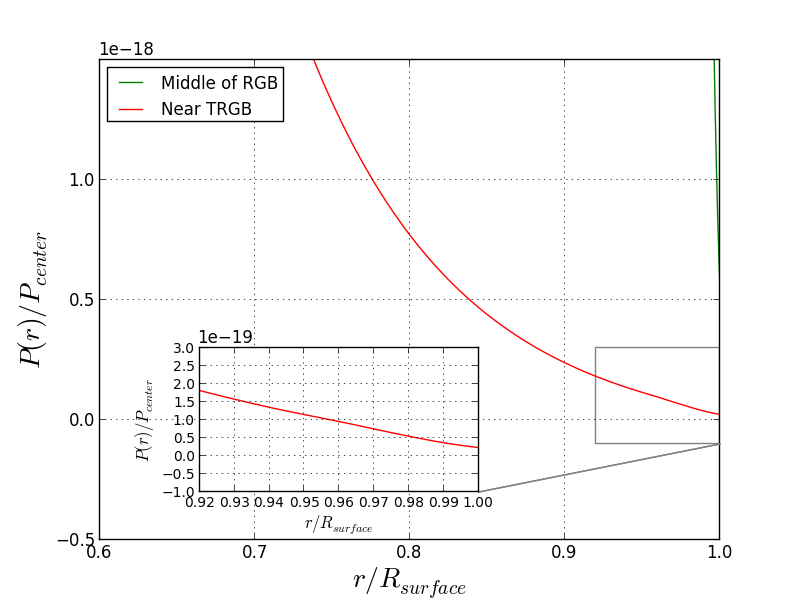}
\hfill     
    \includegraphics[width=0.45\textwidth]{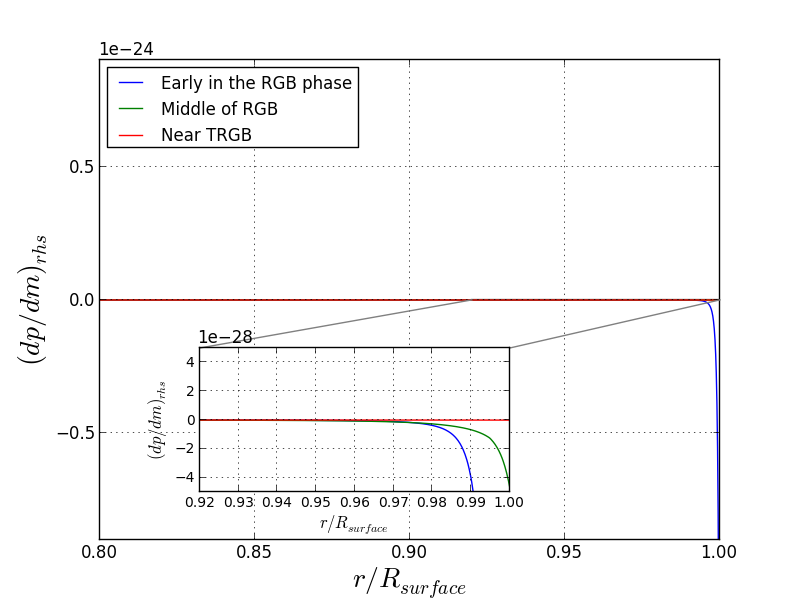}
 \hfill   
 \includegraphics[width=0.45\textwidth]{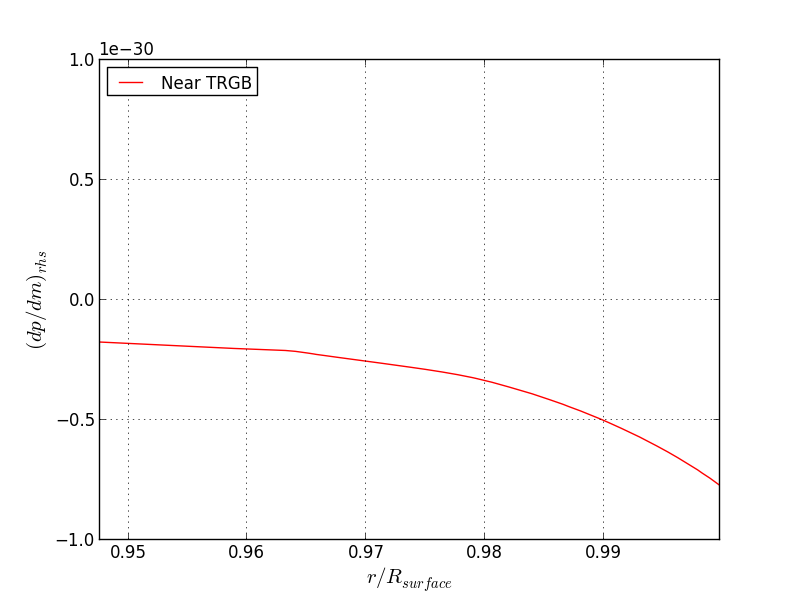}

 \caption{\label{fig:11}The density and pressure of the star are plotted as a function of radius in three epochs during the RGB phase 
for $\beta=\frac{1}{\sqrt{6}}$. bottom row: right hand side of the HSE equation is plotted as a function of radius with 
a closer look at the place of density bump(right panel).}
  \end{figure}

 We have plotted the density and pressure profiles 
 with a closer look at the low density envelope for $\beta=\frac{1}{\sqrt{6}}$ and 
$\beta=800$ in three different epochs during the RGB phase. (1) early in the RGB phase (2) in middle of RGB evolutionary track (3) near TRGB.
In figure~\ref{fig:11}, it can be seen that for $\beta=\frac{1}{\sqrt{6}}$ as the star evolves from main sequence towards TRGB there appears some
 density bumps in the low  density envelope of the star which cannot affect the pressure since for small values of $\beta$,
 the fifth force term in Hydrostatic equilibrium equation of the star is not dominated over the gravitational term in the place of density bump,
 therefore the HSE condition of the star is not violated as it is clear in figure~\ref{fig:11}(bottom plots).

\begin{figure}[tbp]
    \centering
\includegraphics[width=0.45\textwidth]{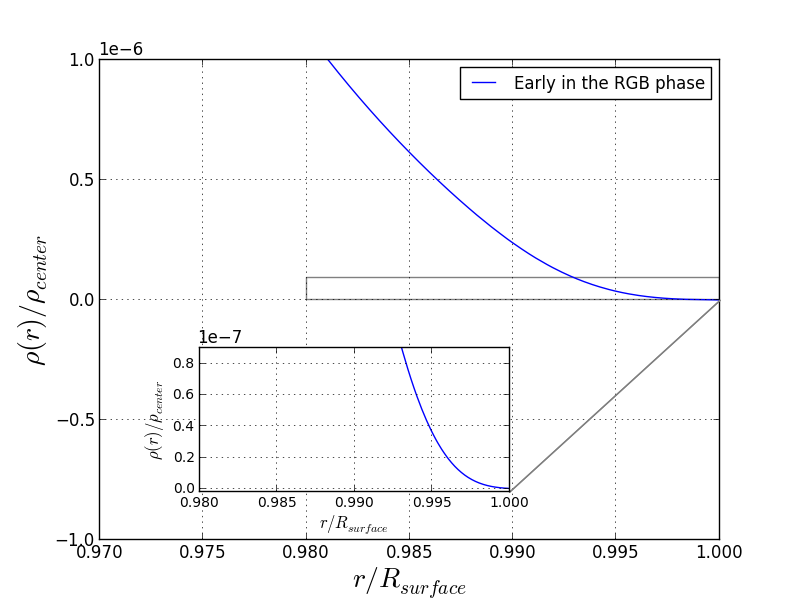}
\hfill
     \includegraphics[width=0.45\textwidth]{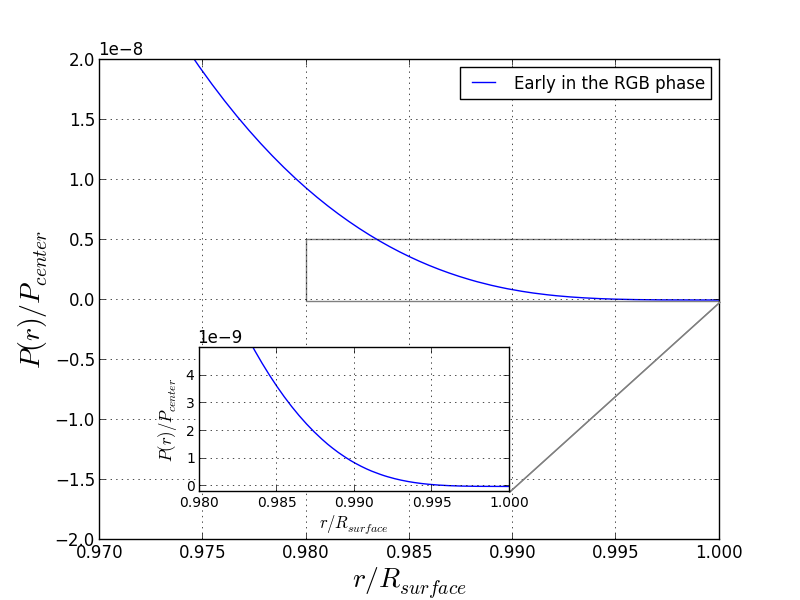}
\hfill     
    \includegraphics[width=0.45\textwidth]{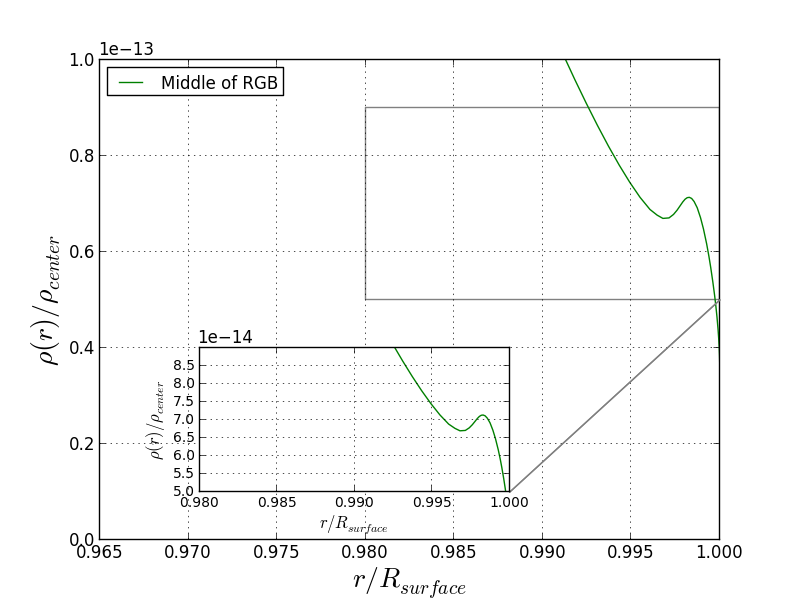}
 \hfill   
 \includegraphics[width=0.45\textwidth]{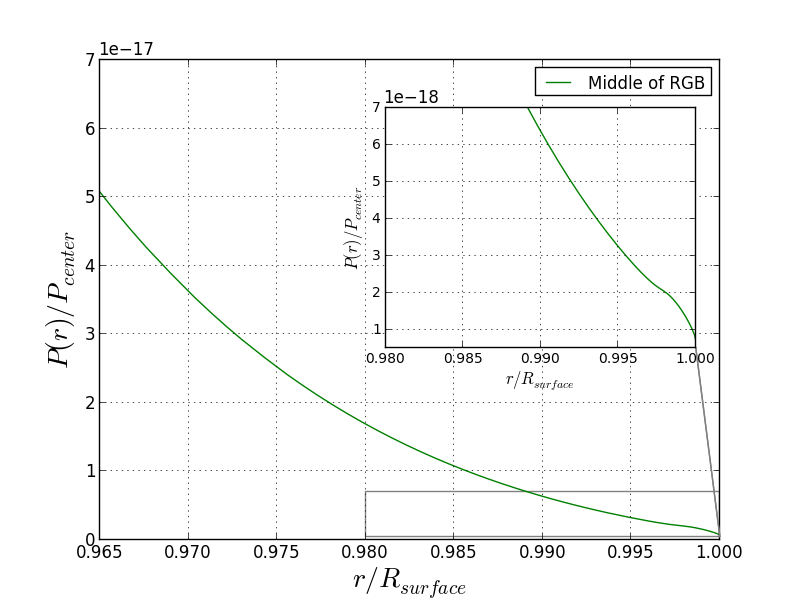}
 \hfill
 \includegraphics[width=0.45\textwidth]{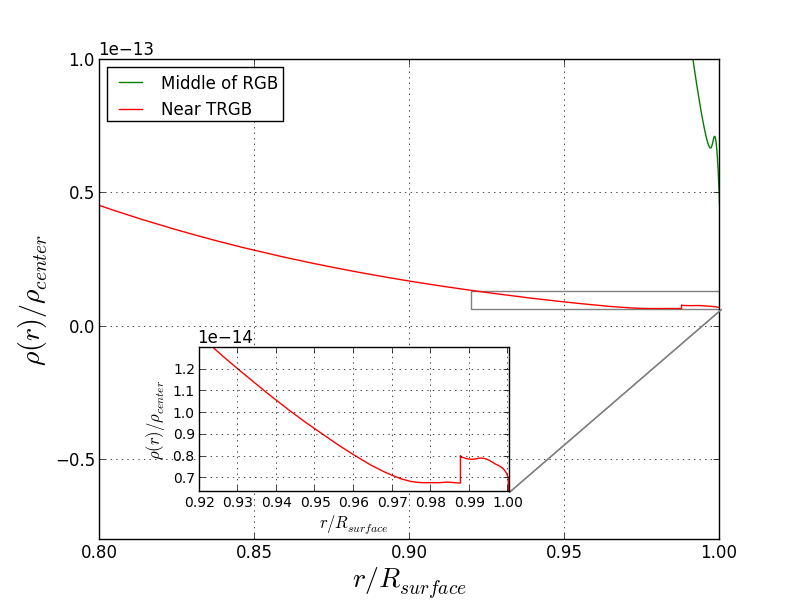}
  \hfill
 \includegraphics[width=0.45\textwidth]{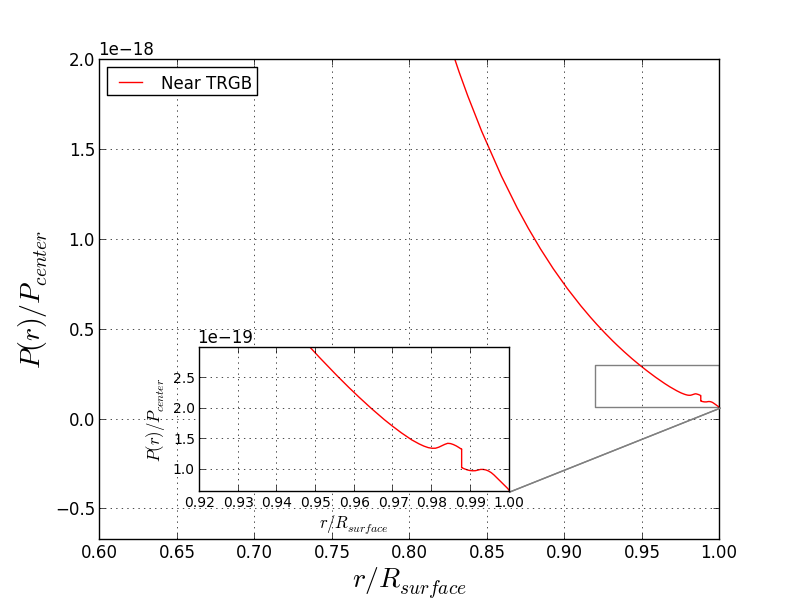}

 \caption{\label{fig:12} The density and pressure of the star are plotted as a function of radius in three epochs during RGB phase for $\beta=800$.
  As it is clear pressure inversion appears near TRGB exactly in the place of density bump. }
  \end{figure}

For high values of coupling constant like $\beta=800$ not only there appears some density bumps but also especially in the place of density bump,
 the fifth force becomes stronger in a way that it can overcome the gravitational force 
in low density regions of envelope which can affect the pressure via HydroStatic 
Equilibrium(HSE) equation of the star(figure~\ref{fig:12}). Since at any radius within the star $\frac{dP}{dr}$ (or $\frac{dP}{dm}$ in MESA code) 
must remain negative to preserve stability~\cite{35}, it makes the  structure unstable.

 This effect is especially powerful in the RGB 
phase because of weak gravitational attraction in low density envelope.

\begin{figure}[tbp]
    \centering
\includegraphics[width=0.49\textwidth]{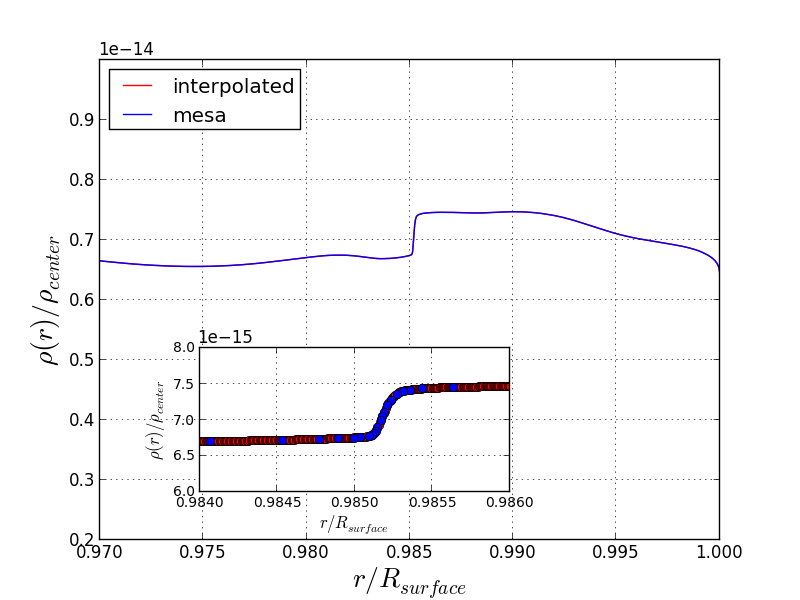}
\hfill
     \includegraphics[width=0.49\textwidth]{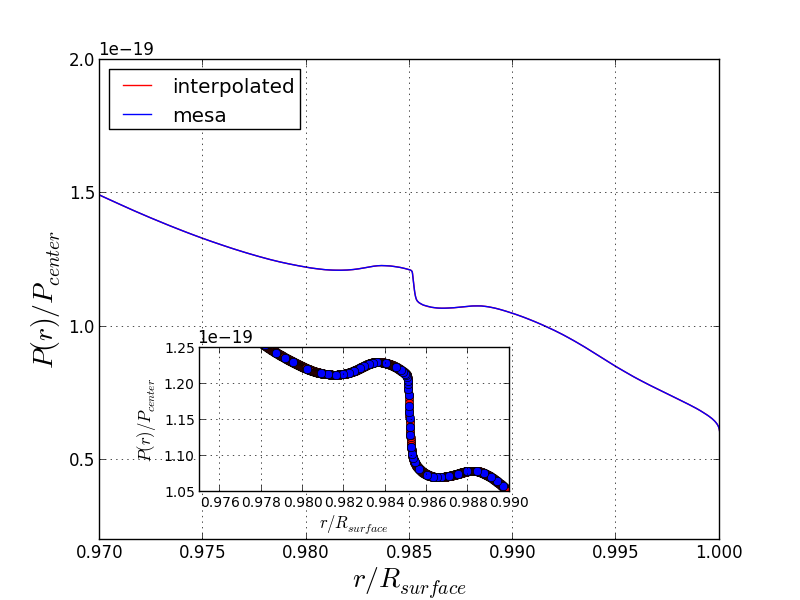}
\hfill     
    \includegraphics[width=0.49\textwidth]{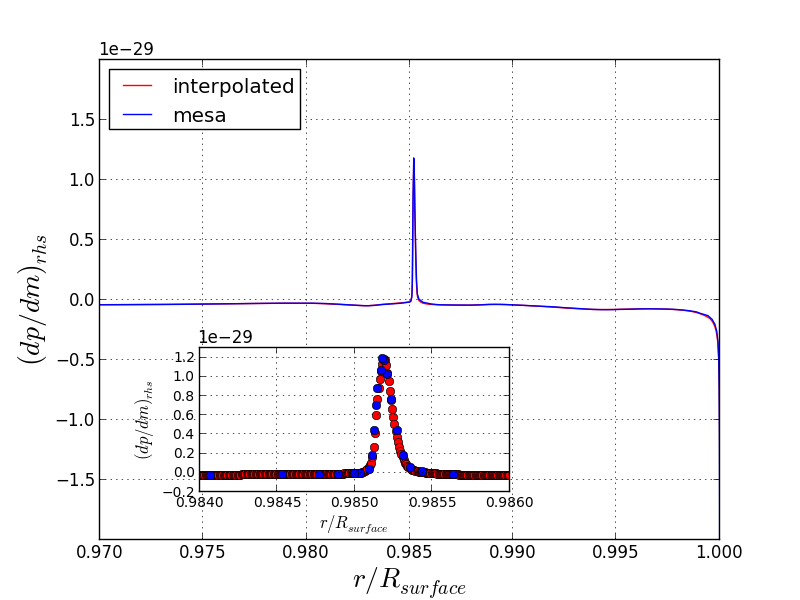}
 \hfill   
 \includegraphics[width=0.49\textwidth]{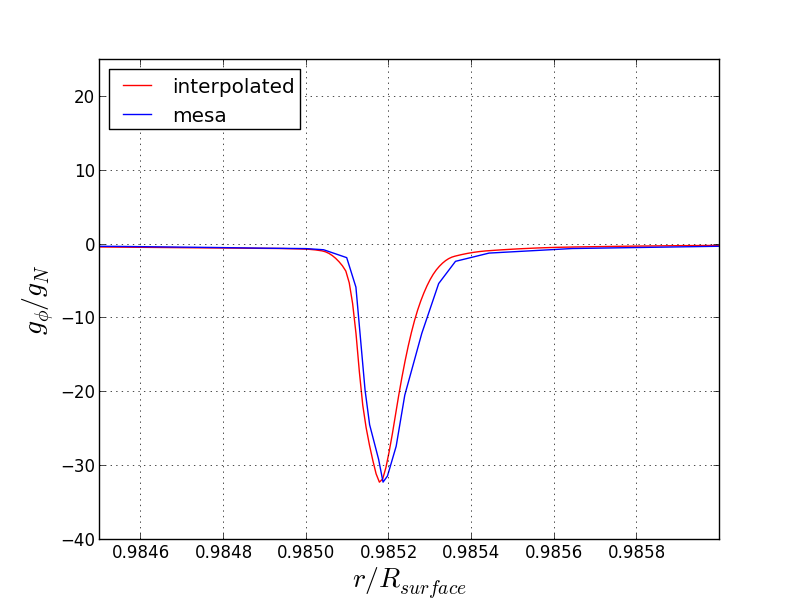}
 \caption{\label{fig:13} The density, pressure, right hand side of HSE and the ratio between fifth force acceleration $g_{\phi}$ to Newtonian acceleration $g_{N}$ of the 
 star near TRGB in the place of bump are plotted as a function of radius for $\beta=500$. It is clear that a sharp pressure gradient must appear exactly in the
 place of density bump which is orders of magnitude bigger than other layers and can have positive sign. }
  \end{figure}

In order to investigate the origin of instability and prove that the instability at the density bump for high values of $\beta$ is physical,
 we have chosen one snapshot of the star for 
$\beta=500$ near TRGB and interpolated the density, radius, mass and pressure to calculate right hand side of HSE equation outside of
 MESA code figure~\ref{fig:13}. The plots show that the sharp field gradient at density bump is real and not numerical.
 In addition, in the place of density bump the fifth force term is repulsive in the HSE equation and is orders of magnitude greater than the Newtonian 
term,
which makes rapid changes in right hand side of HSE equation in MESA code $(\frac{dP}{dm})$, makes it positive and violates the stability condition
 of the star for high value of coupling constant. To sum up, we cannot claim that all the instability appeared in the evolutionary track of the star
 is totally physical, but in fact, it is clear that the stability condition is violated for special values in the parameter space of the chameleon 
scalar field with attractor behaviour, which indicates the existence of physical instability.
%
%
\\

\section{Discussion}

In this paper, we have considered the thin-shell effect and chameleon field profile in inhomogeneous density distributions and shown
 that the fifth force is weaker than what is usually found in homogeneous density 
distributions. The reason is because the scalar field stays closer to the minimum in thin shells near the surface.
Also, the chameleon scalar field evolution inside one solar mass RGB star was studied using Ratra-Peebles potential with index $(n=1)$ and the constant parameter
taken from cosmology. The most important issue was
 to check the validity of quasi-static approximation in the RGB phase. 
To achieve this aim, we have compared the timescale for the oscillation of the scalar field inside the star with the RGB evolution timescale 
using MESA code. 
It was found that the scalar field reaches the effective potential minimum very quickly, and therefore the field gradient is
 negligible during whole RGB phase.  In the end, we checked numerically the stability of the star in the presence of a chameleon scalar 
field  in order to constrain the coupling constant $\beta$.
We have found that since the chameleon fifth force is sensitive to the matter density gradient,
 if there is any density bump in the outer layers of the RGB star, in particular with low densities that the fifth force overcomes the gravitational
 force,
  the bumps can cause an attractive or repulsive fifth force which leads to rapid oscillations that can violate the stability condition of the star
 for high values of $\beta$. That puts an upper limit on the coupling constant order of magnitude about $O(3)$.
Although it would be more accurate to consider modified gravity effects 
in 3D stellar models which were beyond our purpose in this paper.

The main conclusion is that when dealing with scalar fields, like the chameleon, it might be important to pay attention to the structural details.
 Although the constraint 
was found taking into account the stability of RGB star, it would be possible to put tighter constraint on the coupling constant for the chameleon scalar field with attractor behaviour 
 by considering other stellar evolutionary 
phases. Also, it could be interesting to investigate the scalar field behavior and stability of the star using other values in the parameter space of
 the potential.

\acknowledgments
DFM thanks the Research Council of Norway for their support and the NOTUR cluster FRAM.
  This paper is based upon work from COST action CA15117 (CANTATA), supported by COST (European Cooperation in Science and Technology).
We are incredibly grateful to Josiah Schwab, Francis Timmes and MESA community for answering our questions and helpful suggestions.
MTM would like to thank Yashar Akrami, Alessandra Silvestri and Valeri Vardanyan  for useful discussions.

\end{document}